\documentclass[twocolumn,showpacs,preprintnumbers,nofootinbib,amsmath,amssymb]{revtex4}

\usepackage{graphicx}
\usepackage{amsfonts,latexsym}
\usepackage{euscript,amssymb}
\usepackage{epsfig}
\usepackage{dcolumn}
\usepackage{bm}

\catcode`\@=11
\def\Let@{\relax\iffalse{\fi\let\\=\cr\iffalse}\fi}
\def\vspace@{\def\vspace##1{\crcr\noalign{\vskip##1\relax}}}
\def\multilimits@{\bgroup\vspace@\Let@
 \baselineskip\fontdimen10 \scriptfont\tw@
 \advance\baselineskip\fontdimen12 \scriptfont\tw@
 \lineskip\thr@@\fontdimen8 \scriptfont\thr@@
 \lineskiplimit\lineskip
 \vbox\bgroup\ialign\bgroup\hfil$\m@th\scriptstyle{##}$\hfil\crcr}

\def\Sb{_\multilimits@}
\def\endSb{\crcr\egroup\egroup\egroup}
\def\Sp{^\multilimits@}

\newcommand{\ov}{\overline}
\newcommand{\nn}{\nonumber}
\newcommand{\be}[1]{\begin{equation}\label{#1}}
\newcommand{\ee}{\end{equation}}
\newcommand{\ba}[1]{\begin{eqnarray}\label{#1}}
\newcommand{\ea}{\end{eqnarray}}
\newcommand{\rf}[1]{(\ref{#1})}

\renewcommand{\theequation}{\arabic{section}.\arabic{equation}}

\begin{document}

\title{On the problem of inflation in nonlinear multidimensional
cosmological models}

\author{Tamerlan Saidov}\email{tamerlan-saidov@yandex.ru}  \author{Alexander Zhuk}\email{zhuk@paco.net}
\affiliation{Astronomical Observatory and Department of
Theoretical Physics, Odessa National University, 2 Dvoryanskaya
St., Odessa 65082, Ukraine}

%
%
%
%

%
\begin{abstract} We consider a multidimensional cosmological model with nonlinear
quadratic $R^2$ and quartic $R^4$ actions. As a matter source, we
include a monopole form field, D-dimensional bare cosmological
constant and tensions of branes located in fixed points. In the
spirit of the Universal Extra Dimensions models, the Standard
Model fields are not localized on branes but can move in the bulk.
We define conditions which ensure the stable
compactification of the internal space in zero minimum of the effective
potentials. Such effective potentials may have
rather complicated form with a number of local minima, maxima and
saddle points. Then, we investigate inflation in these models. It
is shown that $R^2$ and $R^4$ models can have up to 10 and 22
e-foldings, respectively. These values are not sufficient to solve
the homogeneity and isotropy problem but big enough to explain the recent CMB data. Additionally, $R^4$ model
can provide conditions for eternal topological inflation.
However, the main drawback of the given inflationary models consists in a value of spectral index $n_s$ which is less than observable now $n_s\approx 1$. For example, in the case of $R^4$ model we find $n_s \approx 0.61$.
\end{abstract}

 \maketitle

\vspace{.5cm}

\section{Introduction}
\setcounter{equation}{0}


Recently, very elegant idea of inflation has achieved spectacular
success in explaining the acoustic peak structure seen in CMB (see
e.g. \cite{Bennett}). It is very difficult
to explain correctly the observable large-scale structure formation without taking into
account the stage of early inflation. There is a big number of
different models of inflation. However, the most of them are
poorly related to fundamental physics. In these models, the stage
of inflation occurs due to a special form of scalar field
potential. Here, the origin of scalar field and the form of its
potential is usually remained out of the scope of these papers.
However, it is well known that scalar fields has naturel origin
from higher-dimensional theories. They are geometrical moduli
(radions, gravexcitons) which are related to the shape of internal
spaces (to scale factors of the internal spaces). After
dimensional reduction to four dimensions, scalar field potential
is completely defined by the topology and matter content of
original higher-dimensional model \cite{RZ,GZ1}. Thus, it is of
undoubted interest to realize inflation in these models
(see e.g. \cite{KSS,LW,CKLQ} in string theory and \cite{GZ2} in
multidimensional cosmological models and references therein).

On the other hand, scalar fields with corresponding potentials
originate naturally from nonlinear gravitational models where
Lagrangian is a function of scalar curvature $f(R)$. It is well
known that such models are equivalent to linear ones plus scalar
field. This scalar field corresponds to an additional degree of
freedom of nonlinear models.
For motivation of these theories, see review \cite{SF}. For
example, among others higher-order gravity theories, $f(R)$
theories are free of ghosts and of Ostrogradski instability
\cite{Woodard}. These models attract great attention because
can provide the late-time acceleration of our Universe due to a
special form of scalar field potentials (see e.g.
\cite{SF,NO,GZBR} and references therein), which is interesting
alternative to the cosmological constant. These models can also
provide the stage of early inflation both in four-dimensional (see
the pioneering paper by Starobinsky \cite{Star} and numerous
references in \cite{SF,NO}) and multidimensional
\cite{GZBR,GMZ1,BR} cases.

In our paper we combine both of these approaches. Namely, we
consider multidimensional models with nonlinear action. To start
with, we investigate the most simple linear multidimensional
model. We show that such model can provide power-law inflation.
Unfortunately, it takes place for the branch where internal space
is decompactified \footnote{Each time when we consider
multidimensional cosmological models we should remember about the
problem of the internal space stabilization/compactification. If
such stabilization is absent we confront with the variation of
four-dimensional fundamental constants. General method of the
internal space stabilization was described in \cite{GZ1} and was
applied after that to numerous models. In the present paper, we
follow also this method.}. Then, to get inflation of the external
space with subsequent stabilization of the internal spaces,
we turn to multidimensional nonlinear models with quadratic and
quartic nonlinearities. First, we obtain the conditions of the
internal space compactification (stabilization). Second, for
corresponding effective potentials, we investigate the possibility
of the external space inflation. We show that in the quadratic and
quartic models we can achieve 10 and 22 e-folds, respectively.
These numbers are sufficient to explain the present day CMB date,
but not enough to solve the horizon and flatness problems.
However, 22 e-foldings is rather big number to encourage the
following investigation of the nonlinear multidimensional models
to find theories where this number will approach to 50-60 e-folds. Even
more, this number (50-60) can be reduced in models with long
matter dominated stage followed by inflation with subsequent decay
into radiation. Precisely this scenario takes place for our
models
where we find that e-folds can be reduced by 6 if the mass of
decaying scalar field $m\sim 1$TeV. So, we believe that the number
of e-folds is not a big problem for proposed models. The main
problem consists in spectral index $n_s \approx 0.61$ (for the
quartic model) which is less than observable $n_s \approx 1$. A
possible solution of this problem may consist in more general form
of the nonlinearity $f(R)$. For example, it was observed in
\cite{Ellis} that simultaneous consideration quadratic and quartic
nonlinearities can flatten the effective potential. We postpone the investigation of this problem for our following paper.

To conclude, we want to indicate two interesting features of
models under consideration. First, the quartic model can provide
the topological inflation. Here, due to quantum
fluctuation of scalar fields, inflating domain wall has fractal
structure (inflating domain wall will contain a number of new
inflating domain walls and each such domain walls will contain
again a new inflating walls etc \cite{Linde}). So, we arrive at the eternal inflation. Second, obtained
solution has the property of the self-similarity transformation (see Appendix B). It means that in the case of zero minimum of the effective potential and
fixed positions for extrema in $(\varphi,\phi)-$plane, the change of the hight of extrema results in rescaling of the dynamical characteristics of the model
(graphics of the number of e-folds, scalar fields, the Hubble parameter and the acceleration parameter versus synchronous time) along the time axis. The decrease (increase) of hight in $c$ times ($c$ is a constant) leads  to the stretch (shrink) of these figures along the time axis in $\sqrt{c}$ times.

The paper is structured as follows. In Sec. II we consider the linear (on scalar curvature) model. The nonlinear quadratic $R^2$ and quartic $R^4$ models are investigated in Sec. III and Sec. IY, respectively. Here, we find the range of parameters where the internal space is stabilized and investigate a possibility for the external space inflation. A brief discussion of the obtained results is presented in the concluding Sec. Y.  The Friedmann equations for multi-component scalar field models are reduced to the system of dimensionless first order ODEs in Appendix A. In Appendix B, we show that the dynamical characteristics (e.g. the Hubble parameter, the acceleration parameter) of considered nonlinear models satisfy the self-similarity condition.

\section{Linear model}
\setcounter{equation}{0}


To start with, let us define the topology of our models. We
consider a factorizable $D$-dimensional metric

\be{2.1}
g^{(D)}=  g^{(0)}(x) + L_{Pl}^2 e^{2\beta^1 (x)}g^{(1)}\, ,
\ee
which is defined on a warped product manifold $M = M_0\times M_1$.
$M_0$ describes external $D_0$-dimensional space-time (usually, we
have in mind that $D_0=4$)
and $M_1$ corresponds to $d_1$-dimensional internal space which is
a flat orbifold\footnote{For example, $S^1/Z_2$ and $T^2/Z_2$
which represent circle and square folded onto themselves due to
$Z_2$ symmetry.} with branes in fixed points. Scale factor of the
internal space depends on coordinates $x$ of the external
space-time: $a_1(x) = L_{Pl}e^{\beta^1(x)}$, where $L_{Pl}$ is the
Planck length.

First, we consider the linear model $f(R)=R$ with $D$-dimensional
action of the form
\be{2.2}
S =\frac{1}{2\kappa_D^2}\int_M
d^Dx\sqrt{|g^{(D)}|}\left\{R[g^{(D)}]-2\Lambda_D\right\} +S_m+S_b
\, .
\ee
$\Lambda_D$ is a bare cosmological constant\footnote{Such
cosmological constant can originate from $D$-dimensional form
field which is proportional to the $D$-dimensional world-volume:
$F^{MN\ldots Q}=(C/\sqrt{|g^{(D)}|})\epsilon^{MN\ldots Q}$. In
this case the equations of motion gives $C=const$ and $F^2$ term
in action is reduced to $(1/D!)F_{MN\ldots Q}F^{MN\ldots Q}=
-C^2$.}. In the spirit of Universal Extra Dimension models
\cite{UED}, the Standard Model fields are not localized on the
branes but can move in the bulk. The compactification of the extra
dimensions on orbifolds has a number of very interesting and
useful properties, e.g. breaking (super)symmetry and obtaining
chiral fermions in four dimensions (see e.g. paper by H.-C. Cheng
at al in \cite{UED}). The latter property gives a possibility to
avoid famous no-go theorem of KK models (see e.g. \cite{no-go}).
Additional arguments in favor of UED models are listed in \cite{Kundu}.

Following a generalized Freund-Rubin ansatz \cite{FR} to achieve a
spontaneous compactification $M \rightarrow M = M_0\times M_1$, we
endow the extra dimensions with real-valued solitonic form field
$F^{(1)}$ with an action:
\be{2.3}
S_m=-\frac12 \int_M
d^Dx\sqrt{|g^{(D)}|}\frac{1}{d_1!}\left(F^{(1)}\right)^2\, ,
\ee
This form field is nested in $d_1$-dimensional factor space $M_1$,
i.e. $F^{(1)}$ is proportional to the world-volume of the internal
space. In this case $(1/d_1!)\left(F^{(1)}\right)^2=\bar
f_1^2/a_1^{2d_1}$, where $\bar f_1$ is a constant of integration
\cite{GMZ2}.

Branes in fixed points contribute in action functional \rf{2.2} in
the form \cite{Zhuk}:
\be{2.4}
S_b=\sum_{\phantom{x}^{fixed}_{points}}\left. \int_{M_0} d^4x
\sqrt{ |g^{(0)}(x)}|\; L_b \right|_{\phantom{x}^{fixed}_{point}}\,
,
\ee
where $g^{(0)}(x)$ is induced metric (which for our geometry
\rf{2.1} coincides with the metric of the external space-time  in
the Brans-Dicke frame) and $L_b$ is the matter Lagrangian on the
brane. In what follows, we consider the case where branes are only
characterized by their tensions $L_{b(k)} = -\tau_{(k)}\, ,\,
k=1,2,\ldots ,m$ and $m$ is the number of branes.

Let $\beta_0^1$ be the internal space scale factor at the present
time and $\bar\beta^{1} = \beta^1 - \beta^1_0$ describes
fluctuations around this value. Then, after dimensional reduction
of the action \rf{2.1} and conformal transformation to the
Einstein frame $ g^{(0)}_{\mu \nu } = \left( e^{d_1\bar
\beta^1}\right)^{-2/(D_0-2)} \tilde g^{(0)}_{\mu \nu}$, we arrive
at effective $D_0$-dimensional action of the form
\ba{2.5}
S_{eff} &=& \frac{1}{2\kappa^2_0}\int_{M_0}d^{D_0}x\sqrt{|\tilde
g^{(0)}|}\left\{R[\tilde g^{(0)}]\right.\nn\\&-&\left. \tilde
g^{(0)\mu\nu}\partial_{\mu}\varphi\partial_{\nu}\varphi-2U_{eff}(\varphi )\right\}\, ,
\ea
where scalar field $\varphi$ is defined by the fluctuations of the
internal space scale factor:
\be{2.6}
\varphi \equiv  \sqrt{\frac{d_1(D-2)}{D_0-2}}\; \bar \beta^1\,
\ee
and
$G := \kappa^2_0/8\pi := \kappa^2_D/(8\pi V_{d_1})$ ($V_{d_1}$ is
the internal space volume at the present time) denotes the
$D_0$-dimensional gravitational constant. The effective potential
$U_{eff}(\varphi )$ reads (hereafter we put $D_0=4$):
\ba{2.7}
U_{eff}(\varphi ) &=& e^{-\, \sqrt{\frac{2d_1}{d_1+2}}\;
\varphi}\left[\Lambda_D + f_1^2e^{-2\, \sqrt{\frac{2d_1}{d_1+2}}\;
\varphi}\right.\nn\\&-&\left.\lambda e^{-\, \sqrt{\frac{2d_1}{d_1+2}}\; \varphi}
\right]\, ,
\ea
where $f_1^2 \equiv \kappa^2_D\bar f_1^2/a^{2d_1}_{(0)1}$ and
$\lambda \equiv -\kappa^2_0\sum^m_{k=1}\tau_{(k)} $.

Now, we should investigate this potential from the point of the
external space inflation and the internal space stabilization.
First, we consider the latter problem. It is clear that internal
space is stabilized if $U_{eff}(\varphi )$ has a minimum with
respect to $\varphi$. The position of minimum should correspond to
the present day value $\varphi =0$. Additionally, we can demand
that the value of the effective potential in the minimum position
is equal to the present day dark energy value $U_{eff}(\varphi =0)
\sim \Lambda_{DE} \sim 10^{-57}\mbox{cm}^{-2}$. However, it
results in very flat minimum of the effective potential which in
fact destabilizes the internal space \cite{Zhuk}. To avoid this
problem, we shall consider the case of zero minimum
$U_{eff}(\varphi =0)=0$.

The extremum condition $\left.dU_{eff}/d\varphi \right|_{\varphi
=0} = 0$ and zero minimum condition $U_{eff}(\varphi =0) =0$
result in a system of equations for parameters $\Lambda_D, f_1^2$
and $\lambda$ which has the following solution:
\be{2.8}
\Lambda_D = f_1^2 = \lambda /2\, .
\ee
For the mass of scalar field excitations (gravexcitons/radions) we
obtain: $m^2 = \left.d^2U_{eff}/d \varphi^2\right|_{\varphi =0} =
(4d_1/(d_1+2))\Lambda_D$. In Fig. 1 we present the effective
potential \rf{2.7} in the case $d_1=3$ and $\Lambda_D = 10$. It is
worth of noting that usually scalar fields in the present paper
are dimensionless\footnote{To restore dimension of scalar fields
we should multiply their dimensionless values by
$M_{Pl}/\sqrt{8\pi}$. } and $U_{eff}, \Lambda_D, f_1^2, \lambda$
are measured in $M_{Pl}^2$ units.
\begin{figure*}[htbp]

\centerline{\includegraphics[width=3.5in,height=2in]{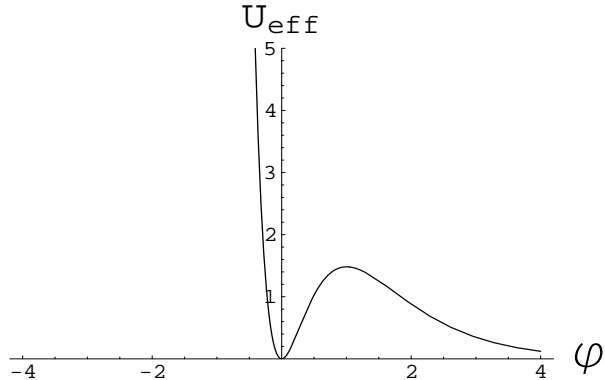}}

\caption {The form of the effective potential \rf{2.7} in the case
$d_1=3$ and  $\Lambda_D = f_1^2 = \lambda /2 =10$.\label{pot}}
\end{figure*}

Let us turn now to the problem of the external space inflation. As
far as the external space corresponds to our Universe, we take
metric $\tilde g^{(0)}$ in the spatially flat
Friedmann-Robertson-Walker form with scale factor $a(t)$. Scalar
field $\varphi $ depends also only on the synchronous/cosmic time
$t$ (in the Einstein frame).

It can be easily seen that for $\varphi
>>0$ (more precisely, for $\varphi
> \varphi_{max} = \sqrt{(d_1+2)/2d_1}\ln 3$) the potential
\rf{2.7} behaves as
\be{2.9}
U_{eff}(\varphi ) \approx \Lambda_D e^{-\sqrt{q}\, \varphi} \, ,
\ee
with
\be{2.10}
q:=\frac{2d_1}{d_1+2}\, .
\ee
It is well known (see e.g.
\cite{RatraPeebles,EM,Ratra1,Stornaiolo}) that for such
exponential potential scale factor has the following asymptotic
form:
\be{2.11}
a(t) \sim t^{2/q}\, .
\ee
Thus, the Universe undergoes the power-law inflation if $q<2$.
Precisely this condition holds for eq. \rf{2.10} if $d_1 \ge 1$.

It can be easily verified that $\varphi
> \varphi_{max}$ is the only region of the
effective potential where inflation takes place. Indeed, in the
region $\varphi <0$ the leading exponents are too large, i.e. the
potential is too steep. The local maximum of the effective
potential $\left.U_{eff}\right|_{max} = (4/27)\Lambda_D$ at
$\varphi_{max} = \sqrt{(d_1+2)/2d_1}\ln 3$ is also too steep for
inflation because the slow-roll parameter $\eta_{max}= \left.
\frac{1}{U_{eff}}\frac{d^2U_{eff}}{d\varphi^2}\right|_{max} =
-\frac{3d_1}{d_1+2}\; \; \Rightarrow \; 1\leq |\eta_{max}|<3$ and
does not satisfy the inflation condition $|\eta |<1$. Topological
inflation is also absent here because the distance between global
minimum and local maximum  $\varphi_{max} = \sqrt{(d_1+2)/2d_1}\ln
3 \leq 1.35$ is less than critical value $\varphi_{cr} \geq 1.65$
(see \cite{Ellis,SSTM,ZhukSaidov}). It is worth of noting that
$\eta_{max}$ and $\varphi_{max}$ depend only on the number of
dimensions $d_1$ of the internal space and do not depend on the
hight of the local maximum (which is proportional to $\Lambda_D$).

Therefore, we have two distinctive regions in this model. In the
first region, at the left of the maximum in the vicinity of the
minimum, scalar field undergoes the damped oscillations. These
oscillations have the form of massive scalar fields in our
Universe (in \cite{GZ1} these excitations were called
gravitational excitons and later (see e.g. \cite{ADMR}) these
geometrical moduli oscillations were also named radions). Their
life-time with respect to the decay $\varphi \to 2\gamma $ into
radiation is \cite{GSZ,Cho,Cho2} $\tau \sim (M_{Pl}/m)^3T_{Pl}$.
For example, we obtain $\tau \sim 10\, \mbox{s} , 10^{-2}\,
\mbox{s}$ for $m \sim 10\, \mbox{TeV} , 10^2\, \mbox{TeV}$
correspondingly. We remind that in our case $m^2 =
(4d_1/(d_1+2))\Lambda_D$. Therefore, this is the graceful exit
region. Here, the internal space scale factor, after decay its
oscillations into radiation, is stabilized at the present day
value and the effective potential vanishes due to zero minimum.
In second region, at the right of the maximum of the potential,
our Universe undergoes the power-low inflation. However, it is
impossible to transit from the region of inflation to the graceful
exit region because given inflationary solution satisfies the
following condition $\dot \varphi >0$. There is also serious
additional problem connected with obtained inflationary solution.
The point is that for the exponential potential of the form
\rf{2.9}, the spectral index reads
\cite{RatraPeebles,Ratra1}\footnote{With respect to conformal
time, solution \rf{2.11} reads $a(\eta) \sim \eta^{1+\beta}$ where
$\beta =-(4-q)/(2-q)$. It was shown in \cite{MSch} that for such
inflationary solution (with $q<2$) the spectral index of density
perturbation is given by $n_s=2\beta +5$ resulting again in
\rf{2.12}.}:
\be{2.12}
n_s=\frac{2-3q}{2-q}\, .
\ee
In our case \rf{2.10}, it results in $n_s = 1-d_1$. Obviously, for
$d_1\ge 1$ this value is very far from observable data $n_s\approx
1$. Therefore, it is necessary to generalize our linear model.

\section{Nonlinear quadratic model}
\setcounter{equation}{0}

As follows from the previous section, we want to generalize the effective potential making it more complicated and having more reach structure.
Introduction of an additional minimal scalar field $\phi$ is one of possible ways.
We can do it "by hand", inserting  minimal scalar field $\phi$ with
a potential $U(\phi )$ in linear action \rf{2.2}\footnote{If such
scalar field is the only matter field in these models, it is
known (see e.g. \cite{GZ2,GMZ1}) that the effective potential can
has only negative minimum. i.e. the models are asymptotical AdS. To
uplift this minimum to nonnegative values, it is necessary to add
form-fields \cite{GMZ2}.}. Then, effective potential takes the
form
\ba{3.1}
U_{eff}(\varphi ,\phi ) &=& e^{-\, \sqrt{\frac{2d_1}{d_1+2}}\;
\varphi}\left[U(\phi ) + f_1^2e^{-2\, \sqrt{\frac{2d_1}{d_1+2}}\;
\varphi}\right. \nn\\ &-&\left.\lambda e^{-\, \sqrt{\frac{2d_1}{d_1+2}}\; \varphi}
\right]\, , \quad
\ea
where we put $\Lambda_D =0$ in \rf{2.2}.

However, it is well known that scalar field $\phi$ can naturally
originate from the nonlinearity of higher-dimensional models
where the Hilbert-Einstein linear lagrangian $R$ is replaced by
nonlinear one $f(R)$. These nonlinear theories are equivalent to
the linear ones with a minimal scalar field (which represents
additional degree of freedom of the original nonlinear theory).
It is not difficult to verify (see e.g. \cite{GMZ1,GMZ2}) that
nonlinear model
\ba{3.2}
S &=&\frac{1}{2\kappa^2_D}\int_M d^Dx\sqrt{|\ov g^{(D)}|}f(\ov R)
\nn\\&-&\frac12 \int_M
d^Dx\sqrt{|g^{(D)}|}\frac{1}{d_1!}\left(F^{(1)}\right)^2
\nn\\ &-&\sum_{k=1}^m \int_{M_0} d^4x \sqrt{ |g^{(0)}(x)}|\; \tau_{(k)}
\ea
is equivalent to a linear one with conformally related metric
\be{3.3}
g_{ab}^{(D)} = e^{2A\phi /(D-2)}\ov g_{ab}^{(D)}
\ee
plus minimal scalar field $\phi =\ln [df/d\ov R\, ]/A$ with a
potential
\be{3.4}
U(\phi )=\frac12 e^{-B\phi }\left[\ov R(\phi )e^{A\phi }-f(\ov
R(\phi) )\right]\, ,
\ee
where $A=\sqrt{(D-2)/(D-1)}=\sqrt{(d_1+2)/d_1+3}$ and
$B=D/\sqrt{(D-2)(D-1)}=A(d_1+4)/(d_1+2)$. After dimensional
reduction of this linear model, we  obtain an effective
$D_0$-dimensional action of the form
\ba{3.5}
S_{eff} &=& \frac{1}{2\kappa^2_0}\int_{M_0}d^{D_0}x\sqrt{|\tilde
g^{(0)}|}\left[R[\tilde g^{(0)}]- \tilde
g^{(0)\mu\nu}\partial_{\mu}\varphi\partial_{\nu}\varphi\right.
\nn\\&-&\left.\tilde
g^{(0)\mu\nu}\partial_{\mu}\phi\partial_{\nu}\phi
-2U_{eff}(\varphi ,\phi )\right]\, ,
\ea
with effective potential exactly of the form \rf{3.1}. It is worth
to note that we suppose that matter fields are coupled to the
metric $g^{(D)}$ of the linear theory (see also analogous approach
in \cite{DSS}). Because in all considered below models both fields
$\varphi$ and $\phi$ are stabilized in the minimum of the
effective potential, such convention results in a simple
redefinition/rescaling of the matter fields and effective
four-dimensional fundamental constants. After such stabilization, the Einstein and Brans-Dicke frames are equivalent each other (metrics $g^{(0)}$ and $\tilde g^{(0)}$ coincide with each other), and linear $g^{(D)}$ and nonlinear $\ov g^{(D)}$ metrics in \rf{3.3} are related via constant prefactor (models became asymptotically linear)\footnote{However, small quantum fluctuations around the minimum of the effective potential distinguish these metrics.}.

Let us consider first the {\bf{quadratic theory}}
\be{3.6}
f(\bar R) = \bar R +\xi \bar R^2 -2\Lambda_D\, .
\ee
For this model the scalar field potential \rf{3.4} reads:
\be{3.7}
U(\phi )=\frac12 e^{-B\phi}\left[\frac{1}{4\xi}\left(e^{A\phi}
-1\right)^2+2\Lambda_D\right]\, .
\ee

It was proven \cite{GZ2} that the internal space is stabilized if
the effective potential \rf{3.1} has a minimum with respect to
both fields $\varphi$ and $\phi$. It can be easily seen from the
form of $U_{eff}(\varphi ,\phi )$ that minimum $\phi_0$
of the potential $U(\phi )$ coincides with the minimum of
$U_{eff}(\varphi ,\phi ) \; : \left.dU/d\phi \right|_{\phi_0} =0
\to \left.\partial_{\phi}U_{eff}\right|_{\phi_0}=0$. For minimum
$U(\phi_0 )$ we obtain \cite{GMZ1}:
\be{3.8}
U(\phi_0) = \frac{1}{8\xi }x_0^{\frac{-D}{D-2}}\left[(x_0-1)^2 +
8\xi\Lambda_D\right]\, ,
\ee
where we denote the constant $x_0 := \exp (A\phi_0)=\left(A-B+\sqrt{A^2+(2A-B)B\,
8\xi\Lambda_D}\, \right)/\left(2A-B\right)$.
It is the global minimum and the only extremum of $U(\phi )$.
Nonnegative minimum of the effective potential $U_{eff}$ takes
place for positive $\xi ,\Lambda_D >0$. If $\xi ,\Lambda_D>0$, the potential $U(\phi)$ has
asymptotic behavior $U(\phi )\to +\infty $ for $\phi \to \pm
\infty$.

The relations \rf{2.8},
where we should make the substitution $\Lambda_D \to U(\phi_0)$,
are the necessary and sufficient conditions of the zero minimum of
the effective potential $U_{eff}(\varphi ,\phi)$ at the point
$(\varphi =0, \phi =\phi_0)$. Thus, if parameters of the quadratic models satisfy the conditions $U(\phi_0)=f_1^2=\lambda/2$, we arrive at zero global minimum: $U_{eff}(0, \phi_0)=0$.

\begin{figure*}[htbp]

\centerline{\includegraphics[width=4.0in,height=2.5in]{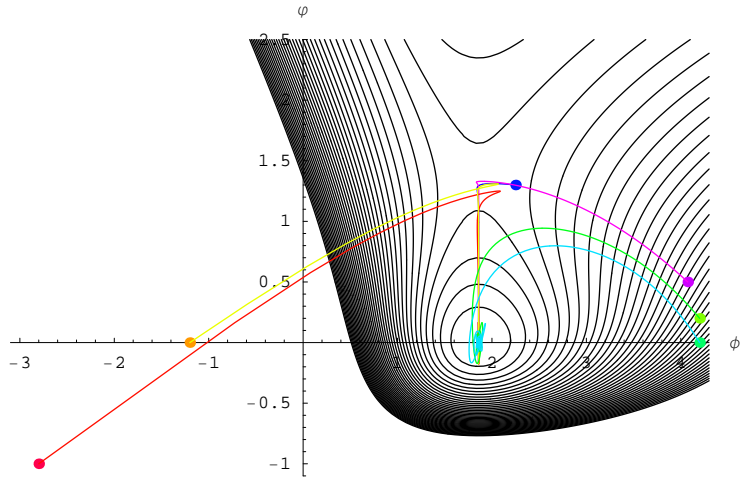}}

\caption {Contour plot of the effective potential $U_{eff}(\varphi
,\phi )$ \rf{3.1} with potential $U(\phi )$ of the form \rf{3.7}
for parameters $d_1=1\, , \xi\Lambda_D  =1 $ and relations
$U(\phi_0) = f_1^2 = \lambda /2$ . This plot clearly shows the
global minimum and the saddle. The colored lines describe
trajectories for scalar fields starting at different initial
conditions. \label{effpotr2}}
\end{figure*}

It is clear that profile $\phi =\phi_0$ of the effective potential
$U_{eff}$ has a local maximum in the region $\varphi >0$ because
$U_{eff}(\varphi ,\phi =\phi_0 ) \to 0$ if $\varphi \to +\infty$.
Such profile has the form shown in Fig.1. Thus, the effective
potential $U_{eff}$ has a saddle point $(\varphi
=\varphi_{max},\phi =\phi_0 )$ where
$\varphi_{max}=\sqrt{(d_1+2)/2d_1}\ln 3$. At this point
$\left.U_{eff}\right|_{max} = (4/27)U(\phi_0)$. The Figure
\ref{effpotr2} demonstrates the typical contour plot of the
effective potential \rf{3.1} with the potential $U(\phi )$ of the
form \rf{3.7} in the vicinity of the global minimum and the saddle
point.

Let us discuss now a possibility of the external space inflation
in this model. It can be easily realized that for all models of
the form \rf{3.1} in the case of local zero minimum at $(\varphi
=0,\phi_0)$, the effective potential will also have a saddle point
at $(\varphi = \varphi_{max}, \, \phi_0)$ with
$\varphi_{max}=\sqrt{(d_1+2)/2d_1}\ln 3 < \varphi_{cr}=1.65$ and
the slow-roll parameter $|\eta_{\varphi}|$ in this point cannot be
less than 1: $|\eta_{\varphi}| =3d_1/(d_1+2)\geq 1$. Therefore,
such saddles are too steep (in the section $\phi =\phi_0$) for the
slow-roll and topological
inflations
However, as we shall see
below, a short period of De Sitter-like inflation is possible if
we start not precisely at the saddle point but first move in the
vicinity of the saddle along the line $\varphi \approx
\varphi_{max}$ with subsequent turn into zero minimum along the
line $\phi \approx \phi_0$. Similar situation happens for
trajectories from different regions of the effective potential
which can reach this saddle and spend here a some time (moving
along the line $\varphi \approx \varphi_{max}$).

Let us consider now regions where the following conditions take
place:
\be{3.9}
U(\phi ) \gg f_1^2e^{-2\, \sqrt{\frac{2d_1}{d_1+2}}\; \varphi}\; ,\;\:
\lambda e^{-\, \sqrt{\frac{2d_1}{d_1+2}}\; \varphi}.
\ee
For the potential \rf{3.7} these regions exist both for negative
and positive $\phi$. In the case of positive $\phi$ with $\exp
(A\phi) \gg \, \mbox{max}\; \left\{1\, ,
(8\xi\Lambda_D)^{1/2}\right\}$ we obtain
\be{3.10}
U_{eff} \approx \frac{1}{8\xi}e^{-\sqrt{q}\;
\varphi}e^{\sqrt{q_1}\; \phi}\, ,
\ee
where $q$ is defined by Eq. \rf{2.10},  $q_1:=(2A-B)^2 =
d_1^2/[(d_1+2)(d_1+3)]$ and $q>q_1$. For potential \rf{3.10} the
slow-roll parameters are\footnote{\label{slowroll}In the case of
$n$ scalar fields $\varphi_i\, (i=1,\ldots ,n)$ with a flat
($\sigma-$model) target space, the slow-roll parameters for the
spatially flat Friedmann Universe read (see e.g. \cite{GZ2,GMZ1}):
$\epsilon \equiv \frac{2}{H^2}\sum_{i=1}^n\left(\partial_i
H\right)^2 \approx \frac{1}{2}|\partial U|^2 / U^2 \, ; \quad
\eta_i \equiv -\ddot \varphi_i / (H\dot\varphi_i)=
2\partial^2_{ii} H /H \approx -\epsilon + \sum_{j=1}^n
\partial^2_{ij}U \partial_j U / (U
\partial_i U)$, where $\partial_i := \partial /\partial \varphi_i$ and $|\partial
U|^2=\sum_{i=1}^n\left(\partial_i U\right)^2$. In some papers (see
e.g. \cite{multi-inflation}) it was introduced a "cumulative"
parameter $\eta \equiv  - \sum_{i=1}^n\ddot \varphi_i\dot\varphi_i
/( H|\dot \varphi|^2) \approx
-\epsilon+\sum_{i,j=1}^n(\partial_{ij}^2 U)(\partial_i
U)(\partial_j U)/(U |\partial U |^2)$ , where $|\dot
\varphi|^2=\sum_{i=1}^n \dot\varphi_i^2$. We can easily find that
for the potential \rf{3.10} parameter $\eta$ coincides exactly
with parameters $\eta_1$ and $\eta_2$.}:
\be{3.11}
\epsilon \approx \eta_1\approx \eta_2 \approx \frac{q}{2}
+\frac{q_1}{2}\,
\ee
and satisfy the slow-roll conditions $\epsilon, \eta_1, \eta_2
<1$. As far as we know, there are no analytic solutions for such
two-scalar-field potential. Anyway, from the form of the potential
\rf{3.10} and condition $q>q_1$ we can get an estimate $a \approx
t^{s}$ with $s\gtrsim 2/q$ (e.g. $2/q = 3,2,5/3$ for $d_1=1,2,3$,
respectively). Thus, in these regions we can get a period of
power-law inflation. In spite of a rude character of these
estimates, we shall see below that external space scale factors
undergo power-law inflation for trajectories passing through these
regions.

Now, we investigate dynamical behavior of scalar fields and the
external space scale factor in more detail. There are no analytic
solutions for considered model. So, we use numerical calculations.
To do it, we apply a Mathematica package proposed in \cite{KP}
adjusting it to our models and notations (see Appendix A).

\begin{figure*}[htbp]

\centerline{\includegraphics[width=3.5in,height=2.5in]{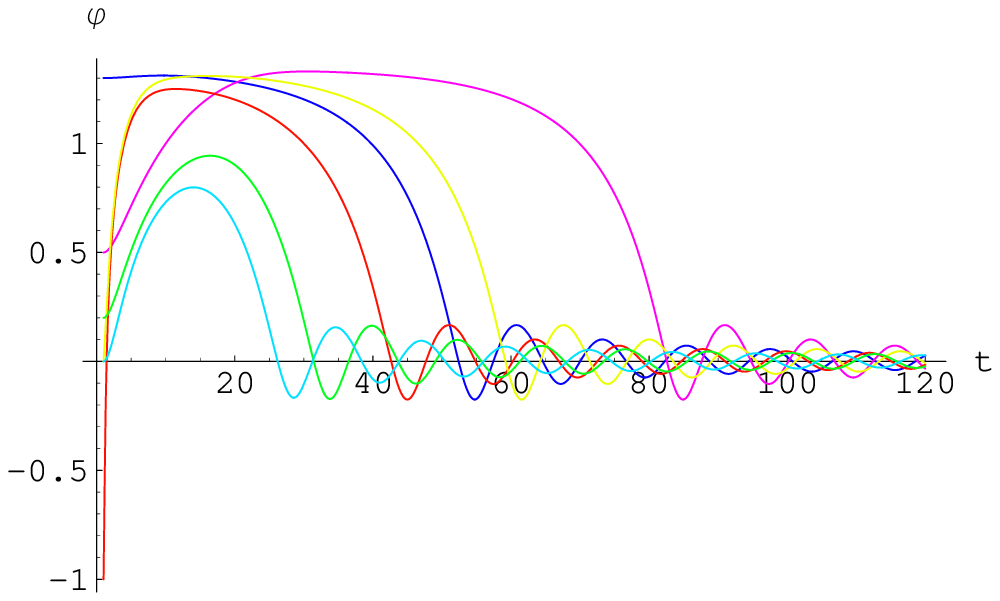}
\includegraphics[width=3.0in,height=2.5in]{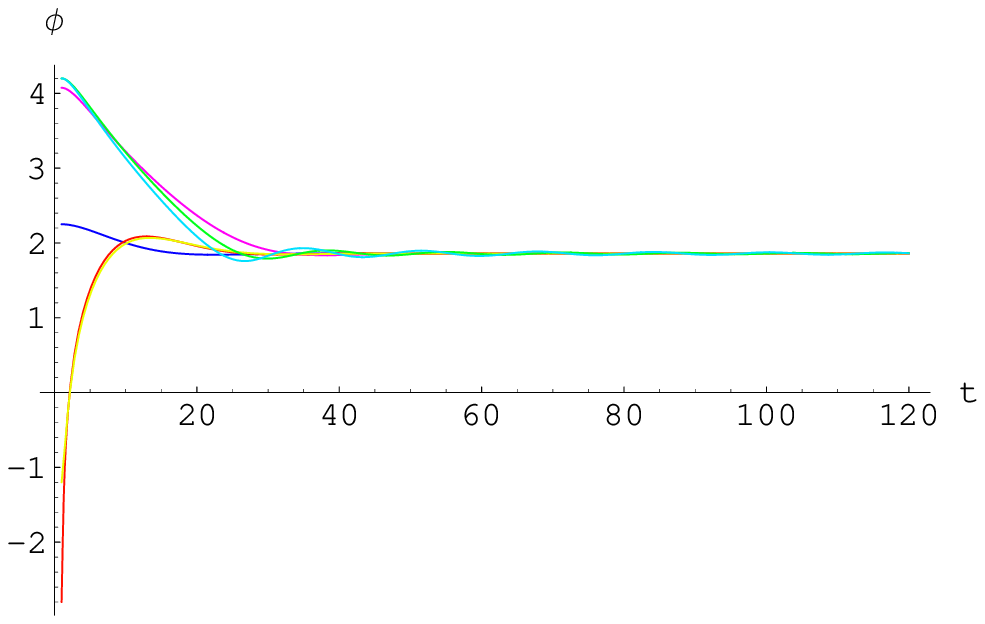}}
\caption {Dynamical behavior of scalar fields $\varphi$ (left
panel) and $\phi$ (right panel) with corresponding initial values
denoted by the colored dots in Fig. \ref{effpotr2}.
\label{fields2}}
\end{figure*}

The colored lines on the contour plot of the effective potential
in Fig. \ref{effpotr2} describe trajectories for scalar fields
$\varphi$ and $\phi$ with different initial values (the colored
dots). The time evolution of these scalar fields\footnote{We
remind that $\varphi$ describes fluctuations of the internal space
scale factor and $\phi$ reflects the additional degree of freedom
of the original nonlinear theory.} is drawn in Fig.
\ref{fields2}. Here, the time $t$ is measured in the Planck times
and classical evolution starts at $t=1$. For given initial
conditions, scalar fields approach the global minimum of the
effective potential along spiral trajectories.

\begin{figure*}[htbp]
\centerline{\includegraphics[width=3.0in,height=2.5in]{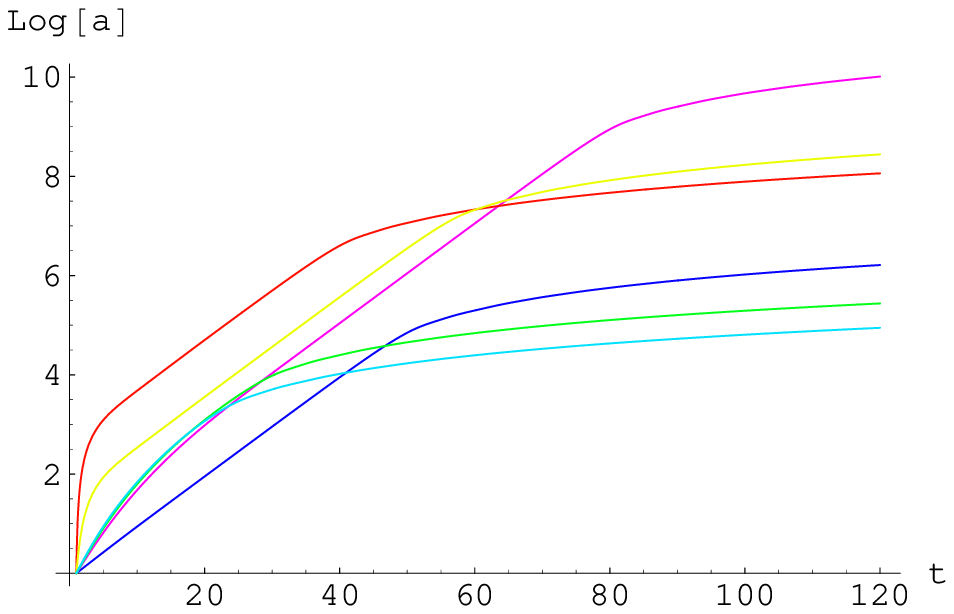}
\includegraphics[width=3.0in,height=2.5in]{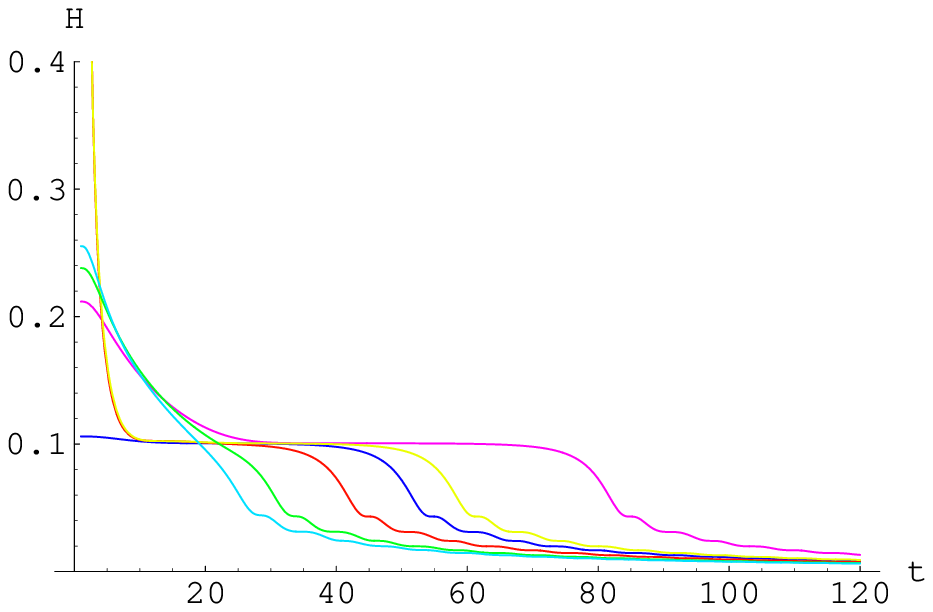}}
\caption {The number of e-folds (left panel) and the Hubble
parameter (right panel) for the corresponding trajectories.
\label{H2}\label{efolds2}}
\end{figure*}

\begin{figure*}[htbp]

\centerline{\includegraphics[width=3.0in,height=2.5in]{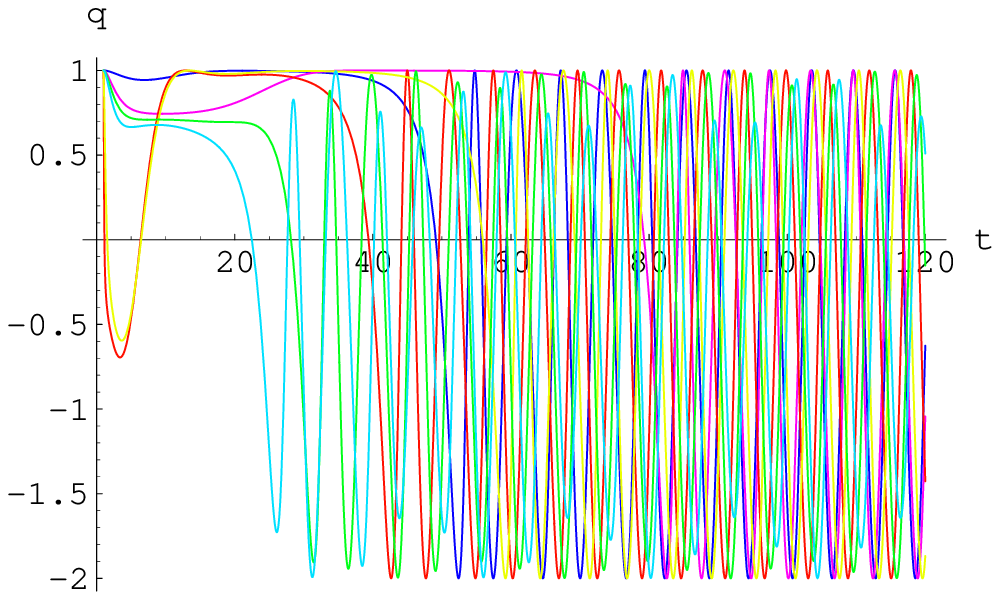}
\includegraphics[width=3.0in,height=2.5in]{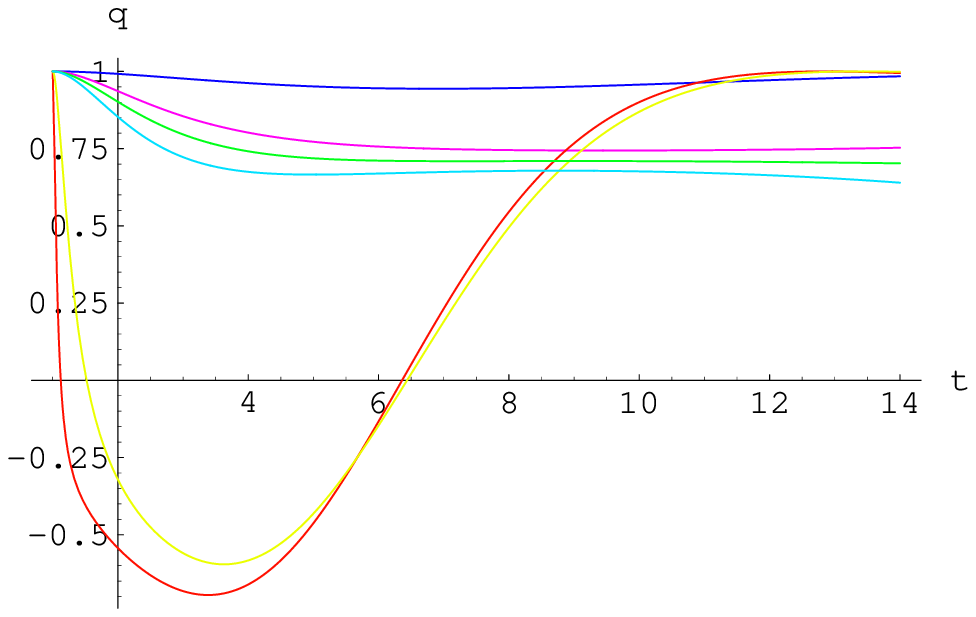}}
\caption {The parameter of acceleration (left panel) and its
magnification for early times (right panel). There are two
different form of acceleration with $q\approx 1$ (De Sitter-like
inflation) and $q\approx 0.75$ (power-law inflation with $s\approx
4$) accordingly. The averaging of $q$ over a few periods of
oscillations results in $\bar q =-0.5$ which corresponds to the
matter dominated decelerating Universe. \label{q2}}
\end{figure*}

We plot in Figure \ref{H2} the evolution of the logarithms of the
scale factor $a(t)$ (left panel) and the evolution of the Hubble
parameter $H(t)$ (right panel) and in Fig. \ref{q2} the evolution
of the parameter of acceleration $q(t)$.

Because for initial condition we use the value $a(t=1) = 1$ (in
the Planck units), then $\log a(t)$ gives the number of e-folds:
$\log a(t) =N(t)$.
The Figure \ref{efolds2} shows that for considered trajectories we
can reach the maximum of e-folds of the order of 10. Clearly, 10
e-folds is not sufficient to solve the horizon and flatness
problems but it can be useful to explain a part of the modern CMB
data. For example, the Universe inflates by $\triangle N \approx 4
$ during the period that wavelengths corresponding to the CMB
multipoles $2\leq l \leq 100$ cross the Hubble radius \cite{CMB}.
However, to have the inflation which is long enough for all modes
which contribute to the CMB to leave the horizon, it is usually
supposed that $\triangle N \ge 15$ \cite{WMAP5}.

The Figure \ref{H2} for the evolution of the Hubble parameter
(right panel) demonstrates that the red, yellow, dark blue and
pink lines have a plateau $H \approx const$. It means that the
scale factor $a(t)$ has a stage of the De Sitter expansion on
these plateaus. Clearly, it happens because these lines reach the
vicinity of the effective potential saddle point and spend there
some time.

The Fig. \ref{q2} for the acceleration parameter defined in \rf{6}
confirms also the above conclusions. According to Eq. \rf{6b},
$q=1$ for the De Sitter-like behavior. Indeed, all these 4 lines
have stages $q\approx 1$ for the same time intervals when $H$ has
a plateau. Additionally, the magnification of this picture at
early times (the right panel of the Figure \ref{q2}) shows that
pink, green and blue lines have also a period of time when $q$ is
approximately constant less than one: $q \approx 0.75$. In
accordance with Eq. \rf{6b}, it means that during this time the
scale factor $a(t)$ undergoes the power-law inflation $a(t)
\propto t^s$ with $s\approx 4$. This result confirms our rude
estimates made above for the trajectories which go through the
regions where the effective potential has the form \rf{3.10}.
After stages of the inflation, the acceleration parameter starts
to oscillate. Averaging $q$ over a few periods of oscillations, we
obtain $\bar q = -0.5$. Therefore, the scale factor behaves as for
the matter dominated Universe: $a(t) \propto t^{2/3}$. Clearly, it
corresponds to the times when the trajectories reach the vicinity
of the effective potential global minimum and start to oscillate
there. It is worth of noting, that there is no need to plot
dynamical behavior for the equation of state parameter $\omega
(t)$ because it is linearly connected with $q$ (see Eq. \rf{6a})
and its behavior can be easily understood from the pictures for
$q(t)$.

As we have seen above for considered quadratic model, the maximal
number of e-folds is near 10. Can we increase this number? To
answer this question, we shall consider a new model with a higher
degree of nonlinearity, i.e. the nonlinear quartic model.


\section{Nonlinear quartic model}
\setcounter{equation}{0}

In this section we consider the nonlinear quartic model
\be{4.1}
f(\bar{R})=\bar{R}+\gamma\bar{R}^{4}-2\Lambda_{D}\, .
\ee
For this model the scalar field potential \rf{3.4} reads
\cite{GZBR}:
\be{4.2}
U(\phi )= \frac12 e^{-B\phi }\left[\frac34 (4\gamma
)^{-1/3}(e^{A\phi }-1)^{4/3} +2\Lambda_D\right]\, .
\ee
Here, the scalar curvature $\bar{R}$ and scalar field $\phi$ are
connected as follows: $e^{A\phi} \equiv f'= 1+4\gamma\bar{R}^{3}
\quad \Leftrightarrow \quad
\bar{R}=\left[(e^{A\phi}-1)/4\gamma\right]^{1/3}\; .$

We are looking for a solution which has a nonnegative minimum of
the effective potential $U_{eff}(\varphi ,\phi )$ \rf{3.1} where
potential $U(\phi)$ is given by Eq. \rf{4.2}. If $\phi_0$
corresponds to this minimum, then, as we mentioned above (see also
\cite{Zhuk}), $U(\phi_0),\lambda $ and $f_1^2$ should be positive.
To get zero minimum of the effective potential, these positive
values should satisfy the relation of the form of \rf{2.8}:
$U(\phi_0) = f_1^2 = \lambda /2$.
Additionally, it is important to note that positiveness of
$U(\phi_0)$ results in positive expression for $\bar R (\phi_0)>0$
\cite{GZBR}.

Eq. \rf{4.2} shows that potential $U(\phi)$ has the following
asymptotes for positive $\gamma$ and $\Lambda_D$\footnote{Negative
values of $\Lambda_D$ and $\gamma$ may lead either to negative
minima, resulting in asymptotically AdS Universe, or to infinitely large negative
values of $U_{eff}$ \cite{GZBR}. In the present paper we want to
avoid both of these possibilities. Therefore, we shall consider
the case of $\Lambda_D ,\gamma >0$. See also footnote
\ref{gamma}.} : $\phi \to -\infty \Longrightarrow U(\phi ) \approx
\frac12e^{-B\phi}\left[\frac34(4\gamma)^{-1/3}+2\Lambda_D\right]
\to +\infty$ and $\phi\to +\infty \Longrightarrow U(\phi) \approx
\frac38(4\gamma)^{-1/3}e^{(-B+4A/3)\phi}\to +0$. For the latter
asymptote we took into account that $-B+4A/3=
(D-8)/3\sqrt{(D-2)(D-1)}<0$ for $D<8$. Obviously, the total number
of dimensions $D=8$ plays the critical role in quartic nonlinear
theories (see \cite{GZBR,BR,ZhukSaidov2}) and investigations for
$D<8, D=8$ and $D>8$ should be performed separately. To make our
paper is not too cumbersome, we consider the case $D<8$ (i.e.
$d_1=1,2,3$), postponing other cases for our following
investigations.

It is worth of noting that for considered signs of parameters, the
effective potential $U_{eff}(\varphi ,\phi)$ \rf{3.1} acquires
negative values when $\phi \to +\infty$ (and $U(\phi) \to 0$). For
example, if $U(\phi_0) = f_1^2 = \lambda /2$ (the case of zero
minimum of the effective potential), the effective potential
$U_{eff}(\varphi , \phi \to \infty) <0$ for $0<e^{-b\varphi}<2$
and the lowest negative asymptotic value
$\left.U_{eff}\right|_{min} \to -(16/27)\lambda$ takes place along
the line $e^{-b\varphi}=4/3$. Therefore, zero minimum of $U_{eff}$
is local\footnote{It is not difficult to show that the thin shell
approximation is valid for considered model and a tunnelling
probability from the zero local minimum to this negative $U_{eff}$
region is negligible.}.

As we mentioned above, extremum positions $\phi_i$ of the
potential $U(\phi )$ coincide with extremum positions of
$U_{eff}(\varphi ,\phi ) \; : \left.dU/d\phi \right|_{\phi_i} =0
\to \left.\partial_{\phi}U_{eff}\right|_{\phi_i}=0$. The condition
of extremum for the potential $U(\phi)$ reads:
\be{4.3}
\frac{d U_{}}{d \phi}=0\;\Longrightarrow \;
\bar{R}^{4}-\frac{(2+d_{1})}{\gamma(4-d_{1})}\bar{R}+2\Lambda_{D}\frac{(4+d_{1})}{\gamma(4-d_{1})}=0
\, .
\ee
For positive $\gamma$ and $\Lambda_D$ this equation has two real
roots:
\begin{equation}\label{4.4}
\bar{R}_{0(1)}=\frac{\Lambda_{D}}{2}\left(-\sqrt{\frac{2(2+d_{1})}{(4-d_{1})k\sqrt{M}}-M}+\sqrt{M}\right)\;
,
\end{equation}
\begin{equation}\label{4.5}
\bar{R}_{0(2)}=\frac{\Lambda_{D}}{2}\left(\sqrt{\frac{2(2+d_{1})}{(4-d_{1})k\sqrt{M}}-M}+\sqrt{M}\right)\;
,
\end{equation}
where we introduced a dimensionless parameter
\be{4.6}
k:=\gamma\Lambda_D^3\, ,
\ee
which is positive for positive $\gamma$ and $\Lambda_D$, and
quantities $M ,\; \omega$ read
\ba{4.7}
M&\equiv&-2^{10/3}\frac{(4+d_{1})}{\omega^{1/3}}-\frac{1}{3\cdot2^{1/3}k}\frac{\omega^{1/3}}{(4-d_{1})}\;,\\
\label{4.8} \omega&\equiv&
k\left[-27(4-d_{1})(2+d_{1})^{2}\right.\nn\\&+&\left.\sqrt{27^{2}(4-d_{1})^{2}(2+d_{1})^{4}-4\cdot24^{3}k(16-d_{1}^2)^{3}}\;
\right].\nn\\
\ea
It can be easily seen that for $k>0$ we get $\omega <0$ and $M\ge
0 $. To have real $\omega $, parameter $k$ should satisfy the
following condition
\be{4.9}
k\leq\frac{27^{2}(4-d_{1})^{2}(2+d_{1})^{4}}{4\cdot24^{3}(16-d_{1}^2)^{3}}\equiv
k_0\, .
\ee
It is not difficult to verify that roots $\bar R_{0(1,2)}$ are
real and positive if $0<k\le k_0$ and they degenerate for $k \to
k_0\, : \quad \bar{R}_{0(1,2)}\to (\Lambda_D/2)\sqrt{M}$. In this
limit the minimum and maximum of $U(\phi )$ merge into an
inflection point. Now, we should define which of these roots
corresponds to minimum of $U(\phi)$ and which to local maximum.
The minimum condition
\be{4.10}
\left.\frac{d^2U(\phi)}{d\phi^2}\right|_{\phi_0} >0\;
\Longrightarrow \;
\gamma\left[(d_{1}+2)-4\gamma\bar{R}_{0}^{3}(4-d_{1})\right]>0
\ee
results in the following inequality\footnote{As we have already
mentioned above, the condition  $U(\phi_0) >0$ leads to the
inequality $\bar R (\phi_0)>0$ \cite{GZBR}. Taking into account the
condition $d_1<4$, we clearly see that inequality $
(d_{1}+2)+4|\gamma|\bar{R}_{0}^{3}(4-d_{1})<0$ for $\gamma <0$
cannot be realized. This is an additional argument in favor of
positive sign of $\gamma$.\label{gamma}}:
\be{4.11} \gamma >0\; : \quad
(d_{1}+2)-4\gamma\bar{R}_{0}^{3}(4-d_{1})>0\, .
\ee
Thus,  the root $\bar R_{0}$ which corresponds to the minimum of
$U(\phi)$ should satisfy the following condition:
\be{4.12}
0< \bar
R_{0}<\left(\frac{d_{1}+2}{4\gamma(4-d_{1})}\right)^{1/3}\, .
\ee
Numerical analysis shows that $\bar{R}_{0(1)}$ satisfies these
conditions and corresponds to the minimum. For $\bar{R}_{0(2)}$ we
obtain that $\bar{R}_{0(2)}
> \left(\frac{d_{1}+2}{4\gamma(4-d_{1})}\right)^{1/3}$ and
corresponds to the local maximum of $U(\phi)$. In what follows we
shall use the notations:
\ba{4.13}
\phi_{min}&=&\frac{1}{A}\ln\left[1+4\gamma
\bar{R}_{0(1)}^{3}\right]\, ,\\
\label{4.14} \phi_{max}&=&\frac{1}{A}\ln\left[1+4\gamma
\bar{R}_{0(2)}^{3}\right]
\ea
and $U(\phi_{min} )\equiv U_{min}\, ,U(\phi_{max} )\equiv U_{max}
$. We should note that $\phi_{min}\, ,\phi_{max}$ and the ratio
$U_{max}/U_{min}$ depend on the combination $k$ \rf{4.6} rather than on $\gamma$ and
$\Lambda_D$ taken separately.

Obviously, because potential $U(\phi )$ has two extrema at
$\phi_{min}$ and $\phi_{max}$, the effective potential
$U_{eff}(\varphi ,\phi)$ may have points of extrema only on the
lines $\phi =\phi_{min}$ and $\phi=\phi_{max}$ where $\partial
U_{eff}/\partial\phi |_{\phi_{min},\phi_{max}}=0$. To find these
extrema of $U_{eff}$, it is necessary to consider the second
extremum condition $\partial U_{eff}/\partial \varphi =0$ on each
line separately:
\be{4.15} \frac{\partial U_{eff}}{\partial \varphi} = 0 \Longrightarrow
\left\{\begin{array}{cc}
-U_{min}-3f_1^{2}\chi^{2}_{1}+2\lambda\chi_{1}=0 \;, \\
\\
-U_{max}-3f_1^{2}\chi^{2}_{2}+2\lambda\chi_{2}=0 \;,  \\
\end{array}\right.
\ee
where $\chi_{1}\equiv \exp \left(-\sqrt{2d_1/(d_1+2)}\, \varphi_{1}\right)>0$ and
$\chi_{2} \equiv\exp \left(-\sqrt{2d_1/(d_1+2)}\, \varphi_{2}\right)>0$;
$\varphi_1$ and $\varphi_2$ denote positions of extrema on
the lines $\phi =\phi_{min}$ and $\phi=\phi_{max}$, respectively.
These equations have solutions
\ba{4.16}
\chi_{1(\pm)}=\alpha&\pm& \sqrt{\alpha^{2}-\beta}\,,\nn\\
\alpha&\geq&\sqrt{\beta}\equiv\alpha_{1}\,;\\
\label{4.17} \chi_{2(\pm)}=\alpha&\pm&
\sqrt{\alpha^{2}-\beta\frac{U_{max}}{U_{min}}}\,,\nn\\
\alpha&\geq&\sqrt{\beta\frac{U_{max}}{U_{min}}}\equiv\alpha_{2}>\alpha_1\,;
\ea
where we have introduced the notations:
$\alpha\equiv\lambda/(3f_1^{2})$ and $\beta\equiv
U_{min}/(3f_1^{2})$. These equations show that there are 5
different possibilities which are listed in the Table
\ref{ttable}.
\begin{table*}
\centering
\caption{The number of extrema of the effective potential
$U_{eff}$ depending on the relation between parameters.
}\label{ttable}

\vspace{0.5cm}
\begin{tabular}{|c|c|c|c|c|}
  \hline
$0<\alpha<\alpha_1 $&$ \alpha=\alpha_1$ &
$\alpha_1<\alpha<\alpha_2$ &$ \alpha=\alpha_2 $&$ \alpha>\alpha_2
$
 \\
\hline
\begin{tabular}{c}
 \textbf{ no extrema}
  \\
  \\
\end{tabular} &
\begin{tabular}{c}
 \textbf{one extremum}
\\
  (point of
  \\
inflection
 on \\the line
 \\ $\phi=\phi_{min}$)
\end{tabular}  &
\begin{tabular}{c}
  \textbf{two extrema}
\\
 (one minimum
 \\and one saddle
 \\ on the line
 \\ $\phi=\phi_{min}$ )
\end{tabular}  &
\begin{tabular}{c}
\textbf{three extrema}
\\
(minimum and
\\
saddle on the line
\\$\phi=\phi_{min}$,
\\
inflection on
\\the
line
\\   $\phi=\phi_{max}$)
\end{tabular}  &
\begin{tabular}{c}
\textbf{ four extrema}
 \\
 (minimum and
\\saddle on
\\the
line
\\ $\phi=\phi_{min}$
\\
maximum and
\\
saddle on
\\the
line
\\ $\phi=\phi_{max}$)
\end{tabular}
\\
\hline\end{tabular}
\end{table*}

To clarify which of solutions \rf{4.16} and \rf{4.17} correspond
to minima of the effective potential (with respect to $\varphi $)
we should consider the minimum condition
\be{4.18}
\left.\frac{\partial^{2}U_{eff}}{\partial^{2}\varphi}\right|_{min}>0\quad
\Longrightarrow \quad U_{extr}+\chi^{2}9f_1^{2}-4\lambda\chi>0\, ,
\ee
where $U_{extr}$ is either $U_{min}$ or $U_{max}$ and $\chi$
denotes either $\chi_1$ or $\chi_2$. Taking into account relations
\rf{4.15}, we obtain
\be{4.19}
\chi^2 3f_1^{2} -\chi \lambda>0 \quad \Longrightarrow\quad
\chi>\frac{\lambda}{3f_1^2}=\alpha .
\end{equation}
Thus, roots $\chi_{1,2(+)}$ define the positions of local minima
of the effective potential with respect to the variable $\varphi$
and $\chi_{1,2(-)}$ correspond to local maxima (in the direction
of $\varphi$).

Now, we fix the minimum $\chi_{1(+)}$ at the point $\varphi =0 $.
It means that in this local minimum the internal space scale
factor is stabilized at the present day value. In this case
\be{4.20}
\left.\chi_{1(+)}\right|_{\varphi=0}=1=\alpha +
\sqrt{\alpha^{2}-\beta}\quad \Longrightarrow\quad
\alpha=\frac{1+\beta}{2}\, .
\ee
Obviously, we can do it only if\footnote{Particular value $\alpha
=1$ corresponds to the case $\alpha=\alpha_1=1$ where the only extremum is the inflection point with $
\chi_{1(-)}=\chi_{1(+)} = \alpha =1$. Here, $\lambda =
U_{min}=3f_1^2$ and  $U_{eff}(\varphi=0, \phi = \phi_{min})=
-\lambda+U_{min}+f_1^2
>0$.} $\alpha <1 \Rightarrow \beta \in
[0,1)$. For $\chi_{1(-)}$ we get:  $\chi_{1(-)}=\beta$.

Additionally, the local minimum of the effective potential at the
point $(\varphi=0, \phi = \phi_{min})$ should play the role of the
nonnegative four-dimensional effective cosmological constant. Thus, we arrive at the
following condition:
\ba{4.21}
\Lambda_{eff}&\equiv& U_{eff}(\varphi=0, \phi = \phi_{min})\nn\\&=&
-\lambda+U_{min}+f_1^2\geq0\nn\\&\Longrightarrow&
-\alpha+\beta+\frac{1}{3}\geq0\, .
\ea
From the latter inequality and equation \rf{4.20} we get
$\beta\in\left[\frac{1}{3},1\right)$. It can be easily seen that
$\beta =1/3 $ (and, correspondingly, $\alpha =2/3$) results in
$\Lambda_{eff}=0$ and we obtain the mentioned above relations:
$U_{min} = f_1^2 = \lambda /2$. In general, it is possible to
demand that $\Lambda_{eff}$ coincides with the present day dark
energy value $10^{-57}\mbox{cm}^{-2}$. However, it leads to very
flat local minimum which means the decompactification of the
internal space \cite{Zhuk}. In what follows, we shall mainly
consider the case of zero $\Lambda_{eff}$ although all obtained
results are trivially generalized to $\Lambda_{eff} =
10^{-57}\mbox{cm}^{-2}$.

Summarizing our results, in the most interesting case of $\alpha
>\alpha_2$ the effective potential has four extrema: local minimum
at $\left(\left.\varphi\right|_{\chi_{1(+)}}=0,\phi_{min}\right)$,
local maximum at $\left(\varphi|_{\chi_{2(-)}},\phi_{max}\right)$
and two saddle-points at
$\left(\varphi|_{\chi_{1(-)}},\phi_{min}\right)$, and
$\left(\left.\varphi\right|_{\chi_{2(+)}},\phi_{max}\right)$ (see
Fig. \ref{effpotr4}).

\begin{figure}[htbp]
\centerline{ \includegraphics[width=2.5in,height=2.5in]{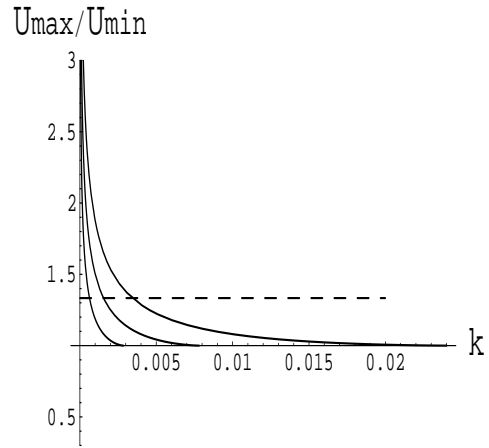}}
\caption{The form of $U_{max}/U_{min}$ as a function of
$k\in(0,k_0]$ for $d_{1}=1,2,3$ from left to right, respectively.
Dashed line corresponds to
 $U_{max}/U_{min}=4/3$. \label{u}}
\end{figure}

\begin{figure*}[htbp]
\centerline{\includegraphics[width=3.5in,height=2.5in]{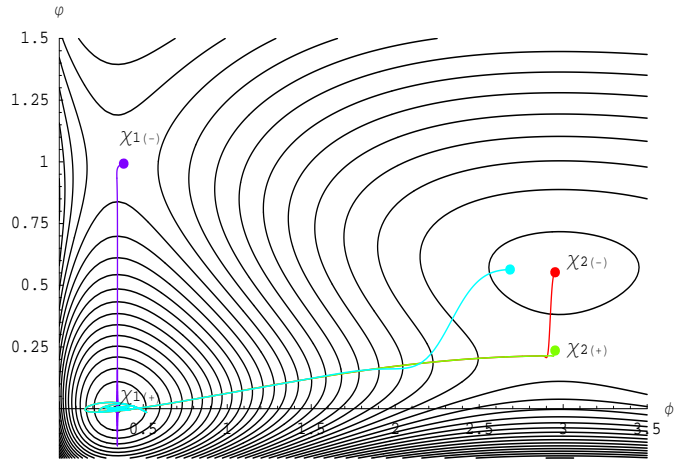}}
\caption {Contour plot of the effective potential $U_{eff}(\varphi
,\phi )$ \rf{3.1} with potential $U(\phi )$ of the form \rf{4.2}
for parameters $\beta = 1/3$, $d_1=3$ and $k=0.004$. This plot
shows the local zero minimum, local maximum and two saddles. The
colored lines describe trajectories for scalar fields starting at
different initial conditions. \label{effpotr4}}
\end{figure*}

We pay particular attention to the case of zero local minimum
$U_{eff}(\left.\varphi\right|_{\chi_{1(+)}} =0,\phi_{min})=0$
where $\beta = 1/3\Longrightarrow \alpha = (1+\beta)/2 =2/3$. To
satisfy the four-extremum condition $\alpha > \alpha_2$, we should
demand
\be{4.22}
\frac{U_{max}}{U_{min}} < \frac43\; .
\ee
The fraction $U_{max}/U_{min}$ is the function of $k$ and depends
parametrically only on the internal space dimension $d_1$.
Inequality \rf{4.22} provides the lower bound on $k$ and numerical
analysis (see Fig. \ref{u})
gives $\tilde{k}(d_{1}=1)\approx0.000625\;;\;\tilde{k}(d_{1}=2)
\approx0.00207\;;\;\tilde{k}(d_{1}=3)\approx0.0035\, .$ Therefore,
effective potentials with zero local minimum will have four
extrema if $k\in (\tilde{k},k_0)$ (where $k_0$ is defined by Eq.
\rf{4.9}). The limit $k \to \tilde k$ results in merging
$\chi_{2(-)} \leftrightarrow \chi_{2(+)}$ and the limit $k\to k_0$
results in merging $\chi_{1(-)} \leftrightarrow \chi_{2(-)}$ and
$\chi_{1(+)} \leftrightarrow \chi_{2(+)}$. Such merging results in
transformation of corresponding extrema into inflection points.
For example, from Fig. \ref{u} follows that $U_{max}/U_{min} \to
1$ for $k \to k_0$.

The typical contour plot of the effective potential with four
extrema in the case of zero local minimum is drawn in Fig.
\ref{effpotr4}. Here, for $d_1=3$ we take $k=0.004 \in (\tilde{k},k_0)$
which gives $\alpha_2 \approx 0.655$. Thus, $\alpha = 2/3 \approx
0.666> \alpha_2$.

Let us investigate now a possibility of inflation for considered
potential. First of all, taking into account the comments in previous section (see a paragraph before Eq. \rf{3.9}), it is clear that topological inflation in the
saddle point $\chi_{1(-)}$ as well as the slow rolling from there
in the direction of the local minimum $\chi_{1(+)}$ are absent. It
is not difficult to verified that the generalized power-low
inflation discussed in the case of the nonlinear quadratic model
is also absent here. Indeed, from Eqs. \rf{3.1} and \rf{4.2} follows
that nonlinear potential $U(\phi)$ can play the leading role in
the region $\phi \to -\infty$ (because $U(\phi)\to 0$ for $\phi
\to +\infty$). In this region $U_{eff} \propto
\exp{(-\sqrt{q}\varphi)}\exp{(-\sqrt{q_2}\phi)}$ where
$q=2d_1/(d_1+2)$ and $q_2=B^2=(d_1+4)^2/[(d_1+2)(d_1+3)]$. For
these values of $q$ and $q_2$ the slow-roll conditions are not
satisfied: $\epsilon \approx \eta_1\approx \eta_2 \approx {q}/{2}
+{q_2}/{2}
>1$. However, there are two promising regions where the stage of
inflation with subsequent stable compactification of the internal
space may take place. We mean the local maximum $\chi_{2(-)}$ and
the saddle $\chi_{2(+)}$ (see Fig. \ref{effpotr4}). Let us estimate the
slow roll parameters for these regions.

\begin{figure*}[htbp]
\centerline{ \includegraphics[width=2.5in,height=2.5in]{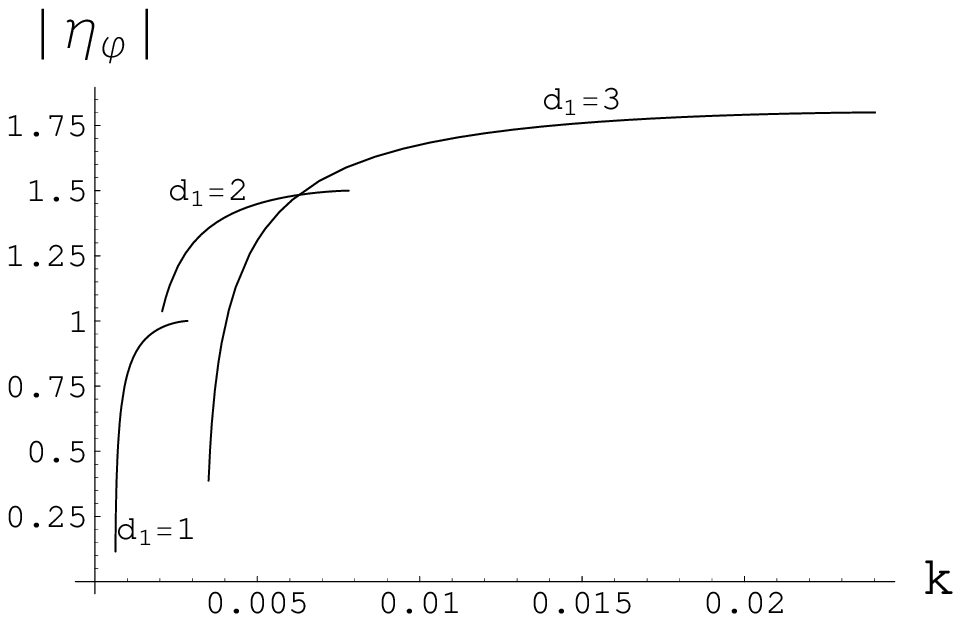}
\includegraphics[width=2.5in,height=2.5in]{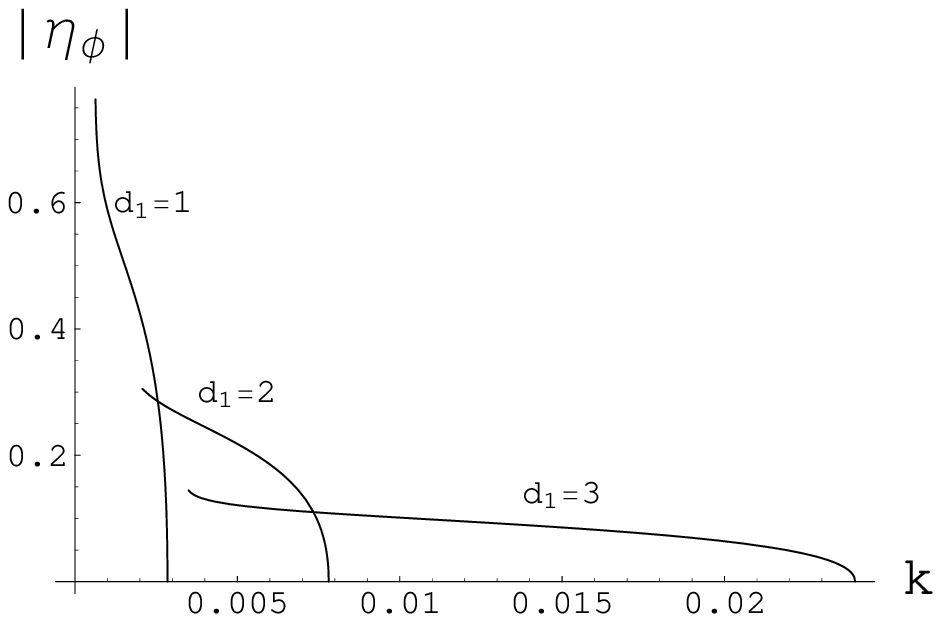}}
\caption{Graphs of $|\eta_{\varphi}|$ (left panel) and
$|\eta_{\phi}|$ (right panel) as functions of $k\in
(\tilde{k},k_0)$ for local maximum $\chi_{2(-)}$ and parameters
$\beta=1/3$ and $d_{1}=1,2,3$.\label{w1}}
\end{figure*}

We consider first the local maximum $\chi_{2(-)}$. It is obvious
that the parameter $\epsilon$ is equal to zero here. Additionally,
from the form of the effective potential \rf{3.1} it is clear that
the mixed second derivatives are also absent in extremum points.
Thus, the slow roll parameters $\eta_1$ and $\eta_2$, defined in
the footnote \rf{slowroll}, coincide exactly with $\eta_{\varphi}$
and $\eta_{\phi}$. In Fig. \ref{w1} we present typical form of
these parameters as functions of $k\in (\tilde k, k_0)$ in the
case $\beta=1/3$ and $d_1=1,2,3$.
These plots show that, for considered parameters, the slow roll
inflation in this region is possible for $d_1=1,3$.

\begin{figure*}[htbp]
\centerline{ \includegraphics[width=2.5in,height=2.5in]{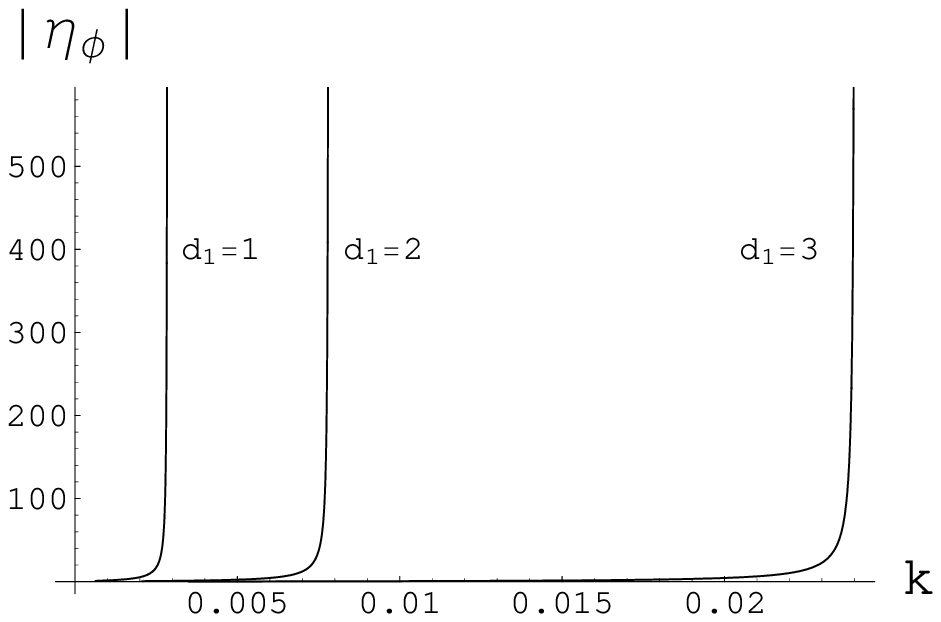}
\includegraphics[width=2.5in,height=2.5in]{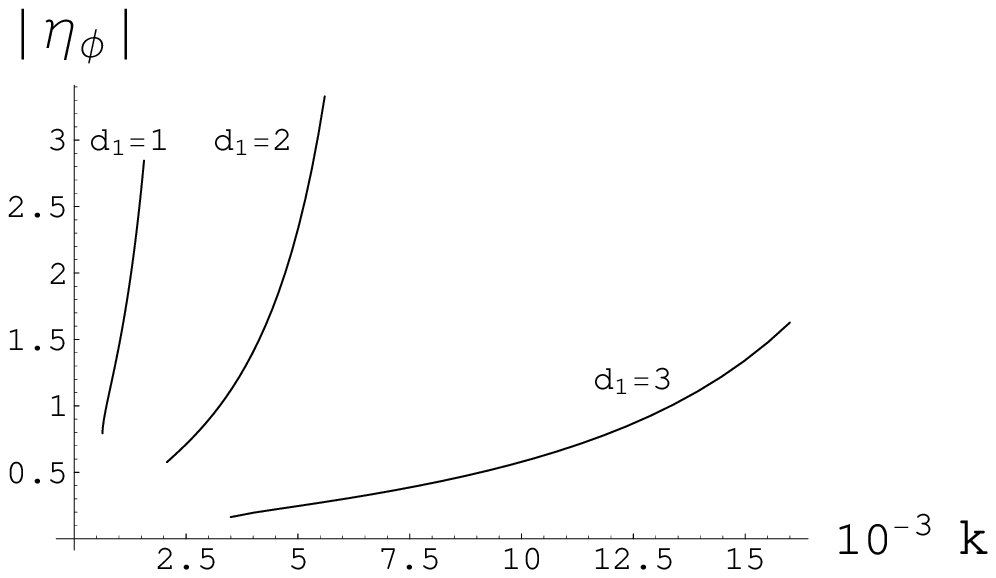}} \caption{Grafs of
$|\eta_{\phi}|$ as functions of $k$ for saddle point $\chi_{2(+)}$
and parameters $\beta=1/3$ and $d_{1}=1,2,3$. Left panel
demonstrates the whole region of variable $k\in (\tilde{k},k_0)$
and right panel shows detailed behavior for small $k$.\label{f1}}
\end{figure*}

The vicinity of the saddle point $\chi_{2(+)}$ is another
promising region. Obviously, if we start from this point, a test
particle will roll mainly along direction of $\phi$. That is why
it makes sense to draw only $|\eta_{\phi}|$. In Fig. \ref{f1} we
plot typical form of $|\eta_{\phi}|$ in the case $\beta=1/3$ and
$d_1=1,2,3$. Left panel represents general behavior for the whole
range of $k\in (\tilde{k},k_0)$ and right panel shows detailed
behavior in the most interesting region of small $k$. It shows
that $d_1=3$ is the most promising case in this region.

\begin{figure*}[htbp]

\centerline{\includegraphics[width=3.5in,height=2.5in]{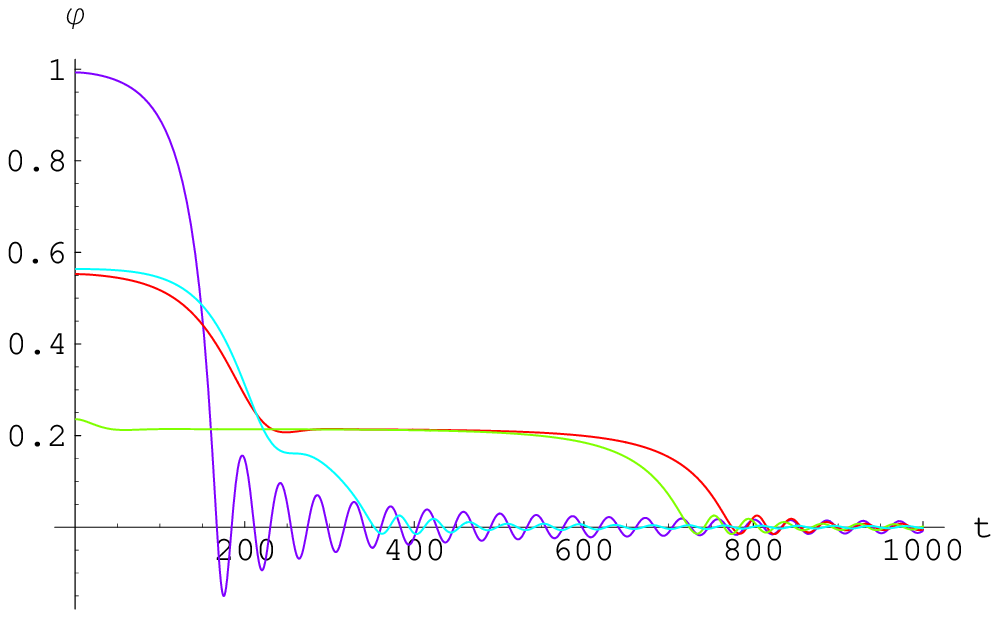}
\includegraphics[width=3.0in,height=2.5in]{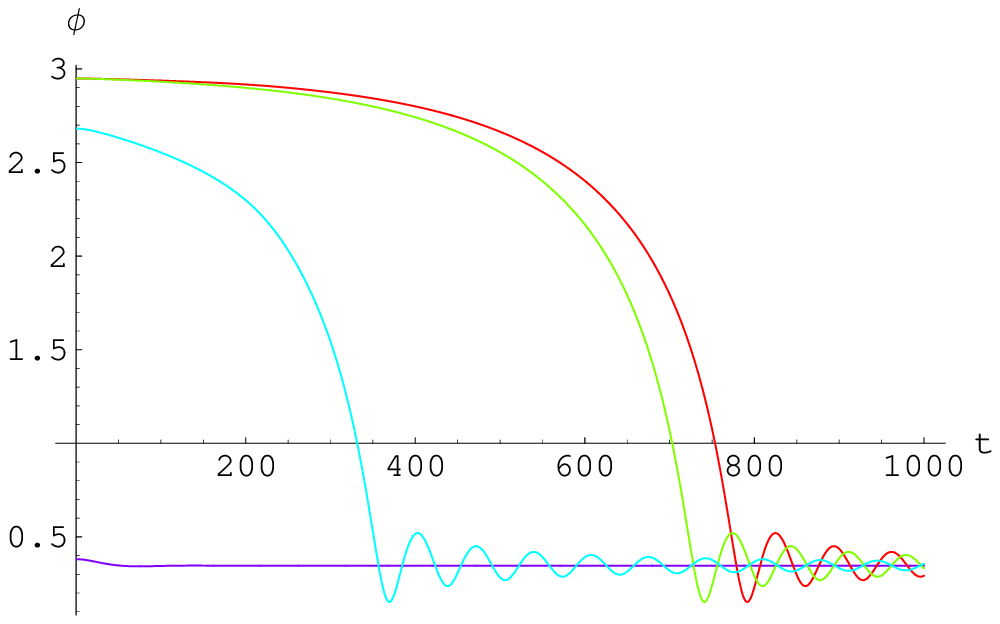}}
\caption {Dynamical behavior of scalar fields $\varphi$ (left
panel) and $\phi$ (right panel) with corresponding initial values
denoted by the colored dots in Fig. \ref{effpotr4}.
\label{fields4}}
\end{figure*}

Now, we investigate numerically the dynamical behavior of scalar
fields and the external space scale factor for trajectories which
start from  the regions $\chi_{1(-)}, \chi_{2(-)}$ and $\chi_{2(+)}$. All
numerical calculations perform for $\beta=1/3, d_1=3$ and
$k=0.004$. The colored lines on the contour plot of the effective
potential in Fig. \ref{effpotr4} describe trajectories for scalar
fields $\varphi$ and $\phi$ with different initial values (the
colored dots) in the vicinity of these extrema points. The time
evolution of these scalar fields is drawn in Fig. \ref{fields4}.
For given initial conditions, scalar fields approach the local
minimum $\chi_{1(+)}$ of the effective potential along the spiral
trajectories.

\begin{figure*}[htbp]
\centerline{\includegraphics[width=3.0in,height=2.5in]{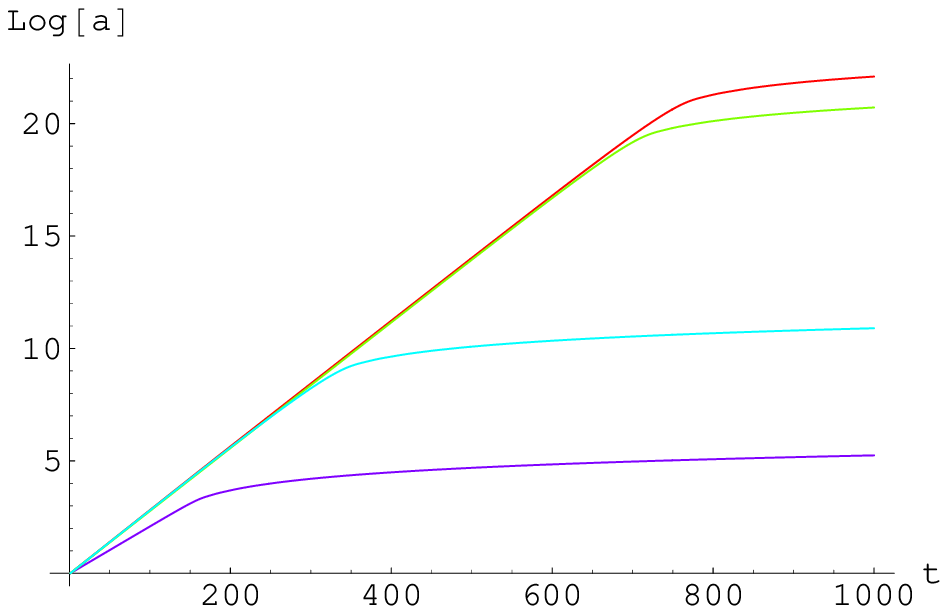}
\includegraphics[width=3.0in,height=2.5in]{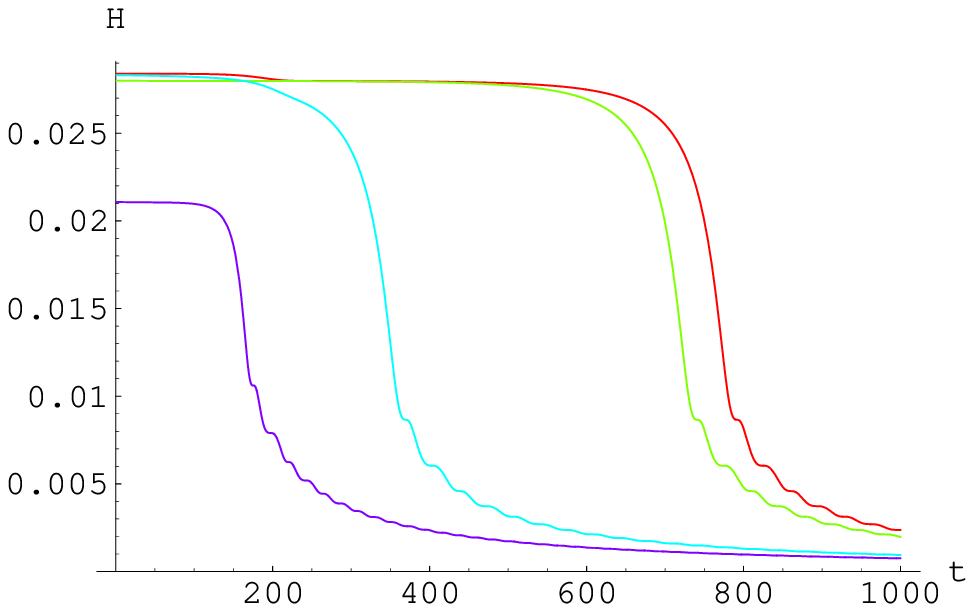}}
\caption {The number of e-folds (left panel) and the Hubble
parameter (right panel) for the corresponding trajectories.
\label{H4}\label{efolds4}}
\end{figure*}

\begin{figure*}[htbp]

\centerline{\includegraphics[width=3.0in,height=2.5in]{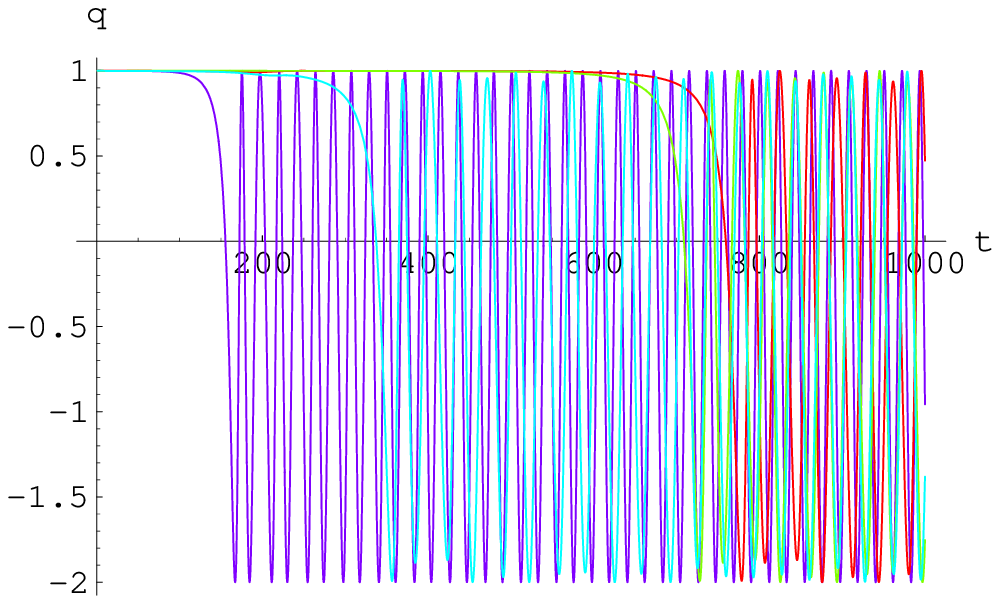}
\includegraphics[width=3.0in,height=2.5in]{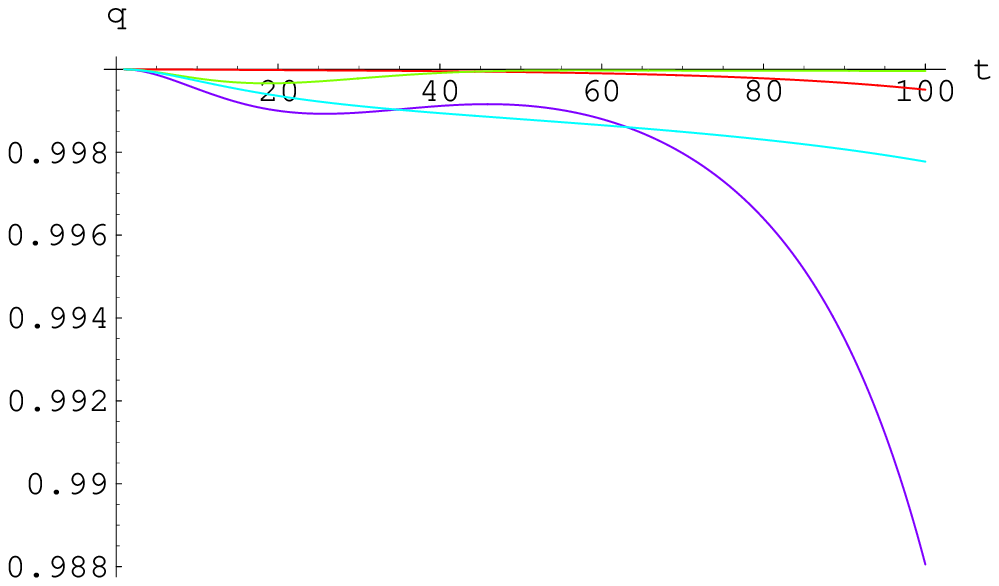}}
\caption {The parameter of acceleration (left panel) and its
magnification for early times (right panel). \label{q4}}
\end{figure*}

We plot in Figure \ref{H4} the evolution of the logarithm of the
scale factor $a(t)$ (left panel) which gives directly the number
of e-folds and the evolution of the Hubble parameter $H(t)$ (right
panel) and in Fig. \ref{q4} the evolution of the parameter of
acceleration $q(t)$.

The Figure \ref{efolds4} shows that for considered trajectories we
can reach the maximum of e-folds of the order of 22 which is long
enough for all modes which contribute to the CMB to leave the
horizon.

The Figure \ref{H4} for the evolution of the Hubble parameter
(right panel) demonstrates that all lines have plateaus $H
\approx const$. However, the red, yellow and blue lines which pass in the vicinity of the saddle $\chi_{2(+)}$ have  bigger
value of the Hubble parameter with respect to the dark blue line which starts from the $\chi_{1(-)}$ region.
Therefore, the scale factor $a(t)$ has stages of the De
Sitter-like expansion corresponding to these plateaus which last approximately
from 100 (dark blue line) up to 800 (red line) Planck times.

The Fig. \ref{q4} for the acceleration parameter confirms also the
above conclusions. All 4 lines have stages $q\approx 1$ for the
same time intervals when $H$ has plateaus. After stages of
inflation, the acceleration parameter starts to oscillate.
Averaging $q$ over a few periods of oscillations, we obtain $\bar
q = -0.5$. Therefore, the scale factor behaves as for the matter
dominated Universe: $a(t) \propto t^{2/3}$. Clearly, it
corresponds to the times when the trajectories reach the vicinity
of the effective potential local minimum $\chi_{1(+)}$ and start to oscillate
there.

Let us investigate now a possibility of the topological inflation \cite{Linde,Vilenkin} if scalar fields
$\varphi, \phi$ stay in the vicinity of the saddle point $\chi_{2(+)}$. As we mentioned in Section 2,
topological inflation in the case of the double-well potential takes place if the distance between a minimum and local maximum bigger than $\Delta\phi_{cr}=1.65$. In this case domain wall is thick enough in comparison with the Hubble radius. The critical ratio of the characteristic thickness of the wall to the horizon scale in local maximum is $r_wH\approx |U/3\partial U_{\phi \phi}|^{1/2} \approx 0.48$ \cite{SSTM} and for topological inflation it is necessary to exceed this critical value. Therefore, we should cheque the saddle $\chi_{2(+)}$ from the point of these criteria.

\begin{figure*}[htbp]
\centerline{
\includegraphics[width=3in,height=2.5in]{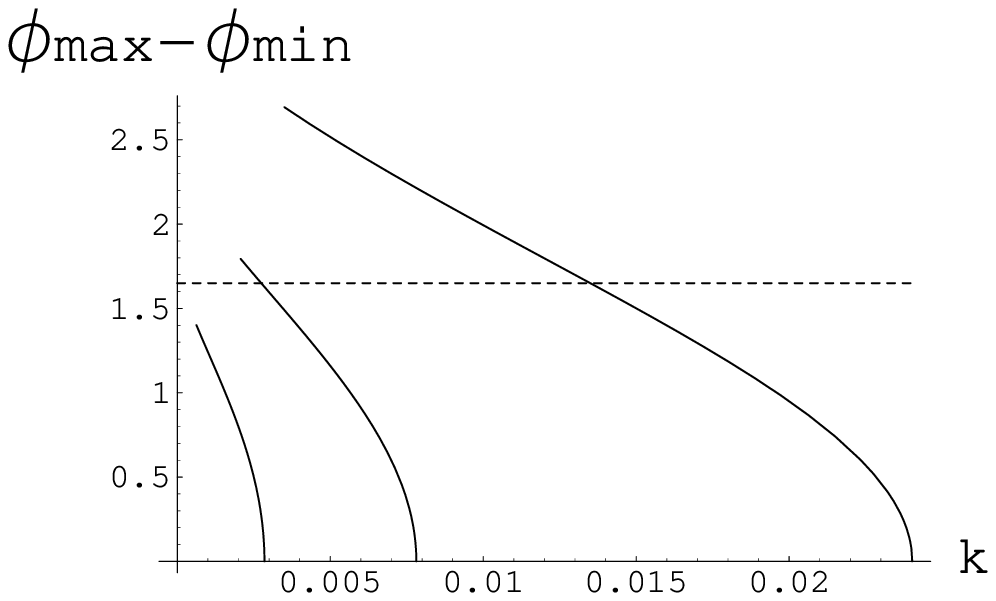}
\includegraphics[width=3in,height=2.5in]{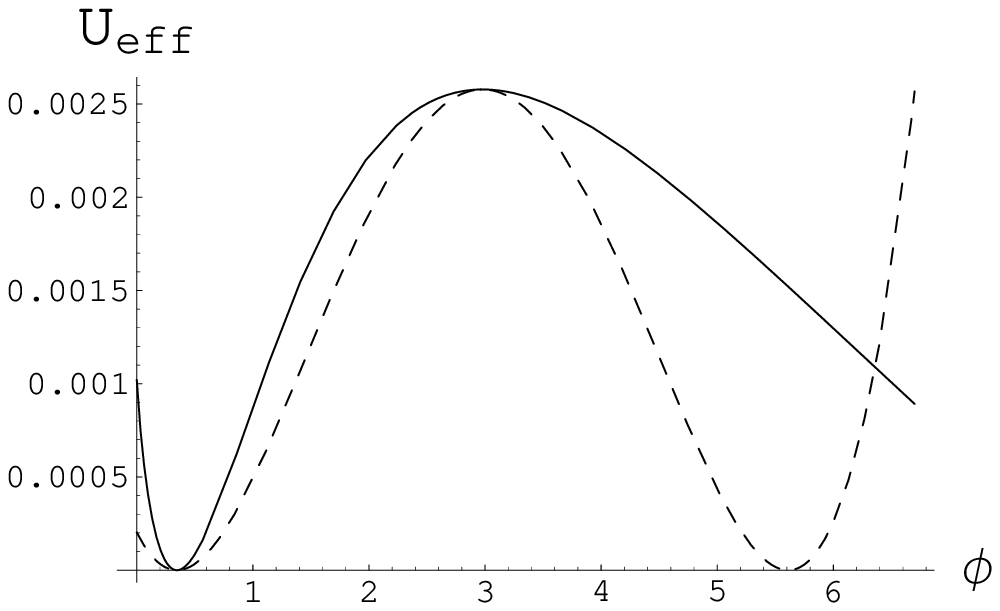}}
 \caption{Left panel demonstrates the difference  $\phi_{max}-\phi_{min}$ (for the profile $\varphi = \varphi|_{\chi_{2(+)}}$)
 as a functions of $k\in (\tilde{k},k_0)$ for parameters
 $\beta=1/3$, and $d_{1}=1,2,3$ (from left to right respectively).
 Dashed line corresponds to $\phi_{max}-\phi_{min}=1.65$. Right panel shows the comparison of the potential $U_{eff}(\varphi|_{\chi_{2(+)}},\phi)$ with a double-well potential for parameters $\beta=1/3, k=0.004$ and $d_1=3$.\label{n}}
\end{figure*}

In Fig. \ref{n} (left panel) we draw the difference  $\Delta \phi =\phi_{max}-\phi_{min}$ for the profile $\varphi = \varphi|_{\chi_{2(+)}}$
 as a functions of $k\in (\tilde{k},k_0)$ in the case
 $\beta=1/3$ for dimensions $d_{1}=1,2,3$. This picture shows that this difference can exceed the critical value if the number of the internal dimensions is $d_1=2$ and $d_1=3$. Right panel of Fig. \ref{n} confirms this conclusion. Here we consider the case $\beta =1/3, \quad k=0.004$ and $d_1=3$. For chosen values of the parameters, $\Delta \phi = 2.63$ which is considerably bigger than the critical value 1.65 and the ratio of the thickness of the wall to the horizon scale is 1.30 which again bigger than the critical value 0.48. Therefore, topological inflation can happen for considered model. Moreover, due to quantum fluctuations of scalar fields, inflating domain wall will have fractal structure: it will contain many other inflating domain walls and each of these domain walls again will contain new inflating domain walls and so on \cite{Linde}. Thus, from this point, such topological inflation is the eternal one.

To conclude this section, we want to draw the attention to one interesting feature of the given model. From above consideration follows that in the case of zero minimum of the effective potential the positions of extrema are fully determined by the parameters $k$ and $d_1$, and for fixed $k$ and $d_1$ do not depend on the choice of $\Lambda_D$. The same takes place for the slow roll parameters. On the other hand, if we keep $k$ and $d_1$, the hight of the effective potential is defined by $\Lambda_D$ (see Appendix B). Therefore, we can change the hight of extrema with the help of $\Lambda_D$ but preserve the conditions of inflation for given $k$ and $d_1$.

However, the dynamical characteristics of the model (drawn in figures \ref{fields4} - \ref{q4}) depend on variations of $\Lambda_D$ by the self-similar manner. It means that the change of hight of the effective potential via transformation $\Lambda_D \to c\Lambda_D$ ($c$ is a constant) with fixed $k$ and $d_1$ results in rescaling of figures \ref{fields4} - \ref{q4} in $1/\sqrt{c}$ times along the time axis.

\section{\label{sec:7}Summary and discussion}
\setcounter{equation}{0}

In our paper we investigated a possibility of inflation in multidimensional cosmological models.
The main attention was paid to nonlinear (in scalar curvature) models with quadratic $R^2$ and quartic $R^4$ lagrangians. These models contain
two scalar fields. One of them corresponds to the scale factor of the internal space and another one is related with the nonlinearity of the original models. The effective four-dimensional potentials in these models are fully determined  by the geometry and matter content of the models. The geometry is defined by the direct product of the Ricci-flat external and internal spaces. As a matter source, we
include a monopole form field, D-dimensional bare cosmological
constant and tensions of branes located in fixed points. The exact form of the effective potentials depends on the relation between parameters of the models and can take rather complicated view with a number of extrema points.

First of all, we found a range of parameters which insures the existence of zero minima of the effective potentials. These minima provide sufficient condition to stabilize the internal space and, consequently, to avoid the problem of the fundamental constant variation. Zero minima correspond to the zero effective four-dimensional cosmological constant. In general, we can also consider positive effective cosmological constant which corresponds to the observable now dark energy. However, it usually requires extreme fine tuning of parameters of models.

Then, for corresponding effective potentials, we investigated the possibility
of the external space inflation. We have shown that for some initial conditions in the quadratic and
quartic models we can achieve up to 10 and 22 e-folds, respectively. An additionally bonus of the considered model is that $R^4$ model can provide conditions for the eternal topological inflation.

Obviously, 10 and 22 e-folds are not sufficient to solve
the homogeneity and isotropy problem but big enough to explain the recent CMB data. To have the inflation which is long enough for modes
which contribute to the CMB, it is usually
supposed that $\triangle N \ge 15$ \cite{WMAP5}. Moreover, 22 e-folds is rather big number to encourage the
following investigations of the nonlinear multidimensional models
to find theories where this number will approach  50-60. We have seen that the increase of the nonlinearity (from quadratic to quartic one) results in the increase of $\triangle N$ in more that two times. So, there is a hope that more complicated nonlinear models can provide necessary 50-60 e-folds.  Besides, this number is reduced in models where long
matter dominated (MD) stage followed inflation can subsequently decay
into radiation \cite{LiddleLyth,BV}. Precisely this scenario takes place for our
models. We have shown for quadratic and quartic nonlinear models, that MD stage with the external scale factor $a\sim t^{2/3}$ takes place after the stage of inflation. It happens when scalar fields start to oscillate near the position of zero minimum of the effective potential. However, scalar fields are not stable. For example, scalar field $\varphi$ decays into two photons $\varphi \to 2 \gamma$ with the decay rate $\Gamma \sim m_{\varphi}^3/M_{Pl}^2$ \cite{GSZ}.
Thus the life time is $\tau_{decay} \sim (M_{Pl}/m_{\varphi})^3t_{Pl}$. The reheating temperature is given by the expression $T_{RH}\sim (m_{\varphi}^3/M_{Pl})^{1/2}$. Therefore, to get $T_{RH} \gtrsim 1$MeV necessary for the nucleosynthesis, we should take $m_{\varphi} \gtrsim 10$TeV.
In paper \cite{BV}, it is shown that for such scenario with intermediate MD stage, the necessary number of e-folds is reduced according to the formula:
\ba{5.1}
\triangle N &=& -\frac{1}{6}\ln \left(\frac{45}{2}g^{-3/2}_{*}\frac{m_{\varphi}^2}{\Gamma M_{Pl}}\right)
\nn\\ &=& -\frac{1}{6}\ln \left(\frac{45}{2}g^{-3/2}_{*}\frac{M_{Pl}}{m_{\varphi}}\right)\, ,
\ea
where $g_{*}$ counts the effective number of relativistic degrees of freedom and we took into account that decaying particles are scalars. This expression weakly depends on $g_{*}$. For example, if $m_{\varphi} \sim 10$TeV we obtain $-6.27 \leq \triangle N \leq -5.11$ for $1 \leq g_{*}\leq 10^2$. Thus, $\triangle N\approx -6$. Therefore, we believe that the number of e-folds is not a big problem for multidimensional nonlinear models. The main problem consists in the spectral index. For example, in the case of $R^4$ model we get $n_s\approx 1+2\eta|_{\chi_{2(+)}}\approx 0.61$ which is less than observable now $n_s\approx 1$. A possible solution of this problem may consist in more general form
of the nonlinearity $f(R)$. It was observed in
\cite{Ellis} that simultaneous consideration quadratic and quartic
nonlinearities can flatten the effective potential and increase $n_s$. We postpone this problem for our following investigations.

\section*{Acknowledgements}
\indent \indent A. Zh. acknowledges the hospitality
of the Theory Division of CERN where this work has been started. A.Zh. would like to
thank the Abdus Salam International Center for Theoretical Physics
(ICTP) for their kind hospitality
during the final stage of this work.This work was supported in part by the
"Cosmomicrophysics" programme of the Physics and Astronomy
Division of the National Academy of Sciences of Ukraine.

\appendix
\section{\label{sec:A}Friedmann equations for multi-component scalar field model}
\renewcommand{\theequation}{A.\arabic{equation}}
\setcounter{equation}{0}

We consider $n$ scalar fields minimally coupled to gravity in four
dimensions.  The effective action of this model reads
\ba{1}
&S& = \frac{1}{16\pi G}\int d^4x \sqrt{|\tilde g^{(0)}|}
\left(R[\tilde g^{(0)}]\right.\nn\\&-&\left. G_{ij}\tilde g^{(0)\mu\nu}
\partial_{\mu}\varphi^i\partial_{\nu}\varphi^j -2 U(\varphi^1,\varphi^2,\ldots)\right)
\ea
where the kinetic term is usually taken in the canonical form:
$G_{ij}=\mbox{diag}(1,1,\ldots )$ (flat $\sigma$ model). Such
multi-component scalar fields originate naturally in
multidimensional cosmological models (with linear or nonlinear
gravitational actions) \cite{GZ1,GZ2,GMZ1}. We use the usual
conventions $c=\hbar =1$, i.e. $L_{Pl}=t_{Pl}=1/M_{Pl}$ and $8\pi
G = 8\pi /M^2_{Pl}$. Here, scalar fields are dimensionless
and potential $U$ has dimension $[U] =
\mbox{lehgth}^{-2}$.

Because we want to investigate dynamical behavior of our Universe
in the presence of scalar fields, we
suppose that scalar fields are homogeneous:
$\varphi^i=\varphi^i(t)$ and four-dimensional metric is
spatially-flat Friedmann-Robertson-Walker one: $\tilde g^{(0)}=
-dt\otimes dt + a^2(t)d\vec{x}\otimes d\vec{x}$.

For energy density and pressure we easily get:
\ba{2}
\rho &=& \frac{1}{8\pi G}\left(\frac12 G_{ij}
\dot\varphi^i\dot\varphi^j +U\right)\,, \nn\\
P&=& \frac{1}{8\pi G}\left(\frac12 G_{ij} \dot\varphi^i\dot\varphi^j - U\right)\,;\\
&\Longrightarrow&
\left\{%
\begin{array}{ll}
  \frac12 G_{ij} \dot\varphi^i\dot\varphi^j= 4\pi
G(\rho+P)\;,\\\\
U= 4\pi G(\rho-P)\;.\label{3}\end{array}%
\right.
\ea

The Friedmann equations  for considered model are
\be{4}
3 \left(\frac{\dot a}{a}\right)^2 \equiv 3 H^2 = 8\pi G \rho =
\frac12 G_{ij} \dot\varphi^i\dot\varphi^j +U\, ,
\ee
and
\be{5}
\dot H = -4\pi G (\rho +P) = - \frac12 G_{ij}
\dot\varphi^i\dot\varphi^j\, .
\ee
From these 2 equations, we obtain the following expression for the
acceleration parameter:
\ba{6}
q &\equiv& \frac{\ddot a}{H^2 a} =  1 -\frac{4\pi G}{H^2} (\rho + P)
= -\frac{8\pi G}{6H^2} (\rho +3P)\nn\\ &=& \frac{1}{6H^2} \left(-4 \times
\frac12 G_{ij} \dot\varphi^i\dot\varphi^j + 2U\right)\, .
\ea
It can be easily seen that the equation of state (EoS) parameter
$\omega = P/\rho$ and parameter $q$ are linearly connected:
\be{6a}
q=-\frac12(1+3\omega)\, .
\ee
From the definition of the acceleration parameter, it follows that
$q$ is constant in the case of the power-law and De Sitter-like
behavior:
\be{6b} q =
\left\{\begin{array}{cc} (s-1)/s \;;\quad
a\propto t^s\, ,\\
1 \;;\quad
a\propto e^{H t}\, . \\
\end{array}\right.
\ee
For example, $q=-0.5$ during the matter dominated (MD) stage where
$s=2/3$.

Because the minisuperspace metric $G_{ij}$ is flat, the scalar
field equations are:
\be{7}
\ddot\varphi^i +3H\dot \varphi^i + G^{ij}\frac{\partial
U}{\partial \varphi^j} = 0\, .
\ee

For the action \rf{1}, the corresponding Hamiltonian is
\be{8a}
\mathcal{H} = \frac{8\pi G}{2a^3}G^{ij}P_iP_j + \frac{a^3}{8\pi G}
U\, ,
\ee
where
\be{8b}
P_{i} = \frac{a^3}{8\pi G} G_{ij} \dot \varphi^j \,
\ee
are the canonical momenta and equations of motion have also the
canonical form
\be{8c}
\dot \varphi^i = \frac{\partial \mathcal{H}}{\partial P_i}\,
,\quad \dot P_i = - \frac{\partial \mathcal{H}}{\partial
\varphi^i}\, .
\ee
It can be easily seen that the latter equation (for $\dot P_i$) is
equivalent to the eq. \rf{7}.

Thus, the Friedmann equations together with the scalar field
equations can be replaced by the system of the first order ODEs:
\ba{11}
 &\dot \varphi^i& = \frac{8\pi G}{a^3} G^{ij}P_j\, ,\\
 &\dot P_i &= - \frac{a^3}{8\pi G}\frac{\partial U}{\partial
\varphi^i}\, ,
\label{12}\\
&\dot a &= a H\, ,\label{13}\\
&\dot H &= \frac{\ddot a}{a} - H^2 \nn\\
&&= \frac16 \left(-4 \times
\frac12 G_{ij} \dot\varphi^i\dot\varphi^j + 2U\right) - H^2
\label{14}\,
\ea
with Eq. \rf{4} considered in the form of the  initial conditions:
\be{16}
H(t=0) = \left.\sqrt{\frac13\left(\frac12 G_{ij}
\dot\varphi^i\dot\varphi^j +U\right)}\; \right|_{t=0}\, .
\ee
We can make these equations dimensionless:
\ba{17}
\frac{d \varphi^i}{M_{Pl}dt} &=& \frac{8\pi }{M^3_{Pl}a^3}
G^{ij}P_j,\nn\\
\Rightarrow  \frac{d \varphi^i}{dt} &=& \frac{8\pi }{a^3} G^{ij}P_j\,;\\
\frac{d P_i}{M_{Pl}dt} &=& - \frac{a^3
M^3_{Pl}}{8\pi}\frac{\partial (U/M^2_{Pl})}{\partial
\varphi^i}\,,\nn\\ \Rightarrow  \frac{d P_i}{dt} &=& -
\frac{a^3}{8\pi}\frac{\partial U}{\partial \varphi^i}\label{18}\,
.
\ea
That is to say the time $t$ is measured in the Planck times
$t_{Pl}$, the scale factor $a$ is measured in the Planck lengths
$L_{Pl}$ and the potential $U$ is measured in the $M^2_{Pl}$
units.

We use this system of dimensionless first order ODEs together with
the initial condition \rf{16} for numerical calculation of the
dynamics of considered models with the help of a Mathematica
package \cite{KP}.


\appendix*
\section*{\label{sec:B}Appendix B:\quad Self-similarity condition}
\renewcommand{\theequation}{B.\arabic{equation}}
\setcounter{equation}{0}

Due to the zero minimum conditions $U(\phi_{min})=f_1^2=\lambda/2$, the effective potential \rf{3.1} can be written in the form:
\ba{1b}
&&U_{eff}(\varphi ,\phi )=U(\phi_{min})e^{-\, \sqrt{\frac{2d_1}{d_1+2}}\;
\varphi}\nn\\
&&\times\left[\frac{U(\phi )}{U(\phi_{min})} + e^{-2\, \sqrt{\frac{2d_1}{d_1+2}}\;
\varphi} -2 e^{-\, \sqrt{\frac{2d_1}{d_1+2}}\; \varphi}
\right]\;.\qquad
\ea
Exact expressions for $U(\phi)$ \rf{3.7} and \rf{4.2} indicate that the ratio
\be{2b}
\frac{U(\phi )}{U(\phi_{min})}=F(\phi,k,d_1)
\ee
depends only on $\phi, k$ and $d_1$. Dimensionless parameter $k=\xi\Lambda_D$ for the quadratic model and $k=\gamma \Lambda_D^3$ for the quartic model. In Eq. \rf{2b} we take into account that $\phi_{min}$ is a function of $k$ and $d_1$: $\phi_{min}=\phi_{min}(k,d_1)$.  Then, $U(\phi_{min})$ defined in Eqs. \rf{3.7} and \rf{4.2} reads:
\be{3b}
U(\phi_{min})=\Lambda_D\tilde F(\phi_{min}(k,d_1),k,d_1)\, .
\ee
Therefore, parameters $k$ and $d_1$ determine fully the shape of the effective potential, and parameter $\Lambda_D$ serves for conformal transformation of this shape.
This conclusion is confirmed also in sections 3 and 4 where we show that
positions of all extrema  depend only on $k$ and $d_1$. Thus, figures \ref{effpotr2}, and  \ref{effpotr4} for contour plots  are defined by $k$ and $d_1$ and will not change with $\Lambda_D$. From the definition of the slow roll parameters it is clear that they also do not depend on the hight of potentials and in our model depend only on $k$ and $d_1$ (see figures \ref{w1} and \ref{f1}). Similar dependence takes place for difference $\Delta \phi =\phi_{max}-\phi_{min}$ drawn in Fig. \ref{n}. Thus the conclusions concerning the slow roll and topological inflations are fully determined by the choice of $k$ and $d_1$ and do not depend on the hight of the effective potential, in other words, on $\Lambda_D$. So, for fixed $k$ and $d_1$ parameter $\Lambda_D$ can be arbitrary. For example, we can take $\Lambda_D$ in such a way that the hight of the saddle point $\chi_{2(+)}$ will correspond to the restriction on the slow roll inflation potential (see e.g. \cite{Lyth}) $U_{eff}\lesssim 2.2 \times 10^{-11}M_{Pl}^4$, or in our notations  $U_{eff}\lesssim 5.5\times 10^{-10}M_{Pl}^2$.

Above, we indicate figures which (for given $k$ and $d_1$) do not depend on the hight of the effective potential (on $\Lambda_D$). What will happen with dynamical characteristics drawn in figures \ref{fields4}, \ref{H4} and \ref{q4} (and analogous ones for the quadratic model) if we, keeping fixed $k$ and $d_1$,  will change $\Lambda_D$? In other words, we keep the positions of the extrema points (in $(\varphi,\phi)$-plane) but change the hight of extrema. We can easily answer this question using the self-similarity condition of the Friedmann equations. Let the potential $U$ in Eqs. \rf{2} and \rf{3} be transformed conformally: $U \to c\, U$ where $c$ is a constant. Next, we can introduce a new time variable $\tau := \sqrt{c}\,  t $. Then, from Eqs. \rf{2}-\rf{5} follows that the Friedmann equations have the same form as for the model with potential $U$ where time $t$ is replaced by time $\tau$. This condition we call the self-similarity. Thus, if in our model we change the parameter $\Lambda_D:\; \Lambda_D\to c\, \Lambda_D$, it results (for fixed $k$ and $d_1$) in rescaling of all dynamical graphics (e.g. Figures \ref{fields4} - \ref{q4}) along the time axis in $1/\sqrt{c}$ times (the decrease of $\Lambda_D$ leads to the stretch of these figures along the time axis and vice versa the increase of $\Lambda_D$ results in the shrink of these graphics). Numerical calculations confirm this conclusion. The property of the conformal transformation of the shape of $U_{eff}$ with change of $\Lambda_D$ for fixed $k$ and $d_1$ can be also called as the self-similarity.


\end{document}